\documentclass[12pt, draftclsnofoot, onecolumn]{IEEEtran}
\renewcommand{\baselinestretch}{1.6}

\usepackage{gensymb}
\usepackage{dsfont}
\usepackage{amsmath,amssymb,amsfonts,amsthm}
\usepackage{epsfig}
\usepackage{cite}
\usepackage{hhline}
\usepackage{multirow}
\usepackage{xcolor}
\usepackage{makecell}
\usepackage{subcaption}
\usepackage{capt-of}
\usepackage[labelformat=simple]{subcaption}

\usepackage{enumerate}
\usepackage{enumitem}
\usepackage{color}
\usepackage{graphicx}
\usepackage{algorithmic}
\usepackage[ruled]{algorithm}
\usepackage{booktabs}
\usepackage{bm}
\usepackage[nolist,printonlyused]{acronym} 
\usepackage{colortbl}

\usepackage[T1]{fontenc}
\usepackage[utf8]{inputenc}
\usepackage{comment} 

\usepackage{pgf}
\usepackage{pgfplots}
\usepackage{tikz}
\usepackage{grffile}
\pgfplotsset{compat=newest}
\usetikzlibrary{plotmarks}
\usepgfplotslibrary{patchplots}
\usetikzlibrary{automata,arrows,positioning,calc,matrix,decorations.markings,decorations.pathreplacing,patterns,arrows.meta}
 \pgfdeclarepatternformonly{south west lines}{\pgfqpoint{-0pt}{-0pt}}{\pgfqpoint{3pt}{3pt}}{\pgfqpoint{3pt}{3pt}}{
        \pgfsetlinewidth{0.4pt}
        \pgfpathmoveto{\pgfqpoint{0pt}{0pt}}
        \pgfpathlineto{\pgfqpoint{3pt}{3pt}}
        \pgfpathmoveto{\pgfqpoint{2.8pt}{-.2pt}}
        \pgfpathlineto{\pgfqpoint{3.2pt}{.2pt}}
        \pgfpathmoveto{\pgfqpoint{-.2pt}{2.8pt}}
        \pgfpathlineto{\pgfqpoint{.2pt}{3.2pt}}
        \pgfusepath{stroke}}
\makeatletter
\newcommand{\vast}{\bBigg@{3}}
\newcommand{\Vast}{\bBigg@{4}}
\makeatother
\renewcommand{\IEEEQED}{\IEEEQEDopen}

\showoutput
\begin{document}

\title{Interference Model Similarity Index and\\ Its Applications to mmWave Networks: Extended version}

\author{Hossein Shokri-Ghadikolaei,~\IEEEmembership{Student Member,~IEEE,} \\ Carlo Fischione,~\IEEEmembership{Member,~IEEE,} and Eytan Modiano,~\IEEEmembership{Fellow,~IEEE} %
\thanks{H. Shokri-Ghadikolaei and C. Fischione are with KTH Royal Institute of Technology, Stockholm,
Sweden (E-mails: \{hshokri, carlofi\}@kth.se).}
\thanks{E. Modiano is with the Laboratory for Information and Decision Systems, Massachusetts Institute of Technology, Cambridge, USA (E-mail: modiano@mit.edu).}
\thanks{A preliminary version of this paper~\cite{Shokri2016OntheAccuracy} has been accepted for presentation at the IEEE International Conference on Communications (ICC), 2016.}
}


\newtheorem{theorem}{Theorem}
\newtheorem{defin}{Definition}
\newtheorem{prop}{Proposition}
\newtheorem{lemma}{Lemma}
\newtheorem{corollary}{Corollary}
\newtheorem{alg}{Algorithm}
\newtheorem{remark}{Remark}
\newtheorem{result}{Result}
\newtheorem{conjecture}{Conjecture}
\newtheorem{example}{Example}
\newtheorem{notations}{Notations}
\newtheorem{assumption}{Assumption}

\newcommand{\be}{\begin{equation}}
\newcommand{\ee}{\end{equation}}
\newcommand{\ba}{\begin{array}}
\newcommand{\ea}{\end{array}}
\newcommand{\bea}{\begin{eqnarray}}
\newcommand{\eea}{\end{eqnarray}}
\newcommand{\netsize}[3]{#1x#2~#3$^2$}
\newcommand{\combin}[2]{\ensuremath{ \left( \ba{c} #1 \\ #2 \ea \right) }}
\newcommand{\diag}{{\mbox{diag}}}
\newcommand{\rank}{{\mbox{rank}}}
\newcommand{\dom}{{\mbox{dom{\color{white!100!black}.}}}}
\newcommand{\range}{{\mbox{range{\color{white!100!black}.}}}}
\newcommand{\image}{{\mbox{image{\color{white!100!black}.}}}}
\newcommand{\herm}{^{\mbox{\scriptsize H}}}  
\newcommand{\sherm}{^{\mbox{\tiny H}}}       
\newcommand{\tran}{^{\mbox{\scriptsize T}}}  
\newcommand{\tranIn}{^{\mbox{-\scriptsize T}}}  
\newcommand{\card}{{\mbox{\textbf{card}}}}
\newcommand{\asign}{{\mbox{$\colon\hspace{-2mm}=\hspace{1mm}$}}}
\newcommand{\ssum}[1]{\mathop{ \textstyle{\sum}}_{#1}}

\newcommand{\vbar}{\raisebox{.17ex}{\rule{.04em}{1.35ex}}}
\newcommand{\vbarind}{\raisebox{.01ex}{\rule{.04em}{1.1ex}}}
\newcommand{\D}{\ifmmode {\rm I}\hspace{-.2em}{\rm D} \else ${\rm I}\hspace{-.2em}{\rm D}$ \fi}
\newcommand{\T}{\ifmmode {\rm I}\hspace{-.2em}{\rm T} \else ${\rm I}\hspace{-.2em}{\rm T}$ \fi}
\newcommand{\B}{\ifmmode {\rm I}\hspace{-.2em}{\rm B} \else \mbox{${\rm I}\hspace{-.2em}{\rm B}$} \fi}
\newcommand{\Hil}{\ifmmode {\rm I}\hspace{-.2em}{\rm H} \else \mbox{${\rm I}\hspace{-.2em}{\rm H}$} \fi}
\newcommand{\C}{\ifmmode \hspace{.2em}\vbar\hspace{-.31em}{\rm C} \else \mbox{$\hspace{.2em}\vbar\hspace{-.31em}{\rm C}$} \fi}
\newcommand{\Cind}{\ifmmode \hspace{.2em}\vbarind\hspace{-.25em}{\rm C} \else \mbox{$\hspace{.2em}\vbarind\hspace{-.25em}{\rm C}$} \fi}
\newcommand{\Q}{\ifmmode \hspace{.2em}\vbar\hspace{-.31em}{\rm Q} \else \mbox{$\hspace{.2em}\vbar\hspace{-.31em}{\rm Q}$} \fi}
\newcommand{\Z}{\ifmmode {\rm Z}\hspace{-.28em}{\rm Z} \else ${\rm Z}\hspace{-.38em}{\rm Z}$ \fi}

\newcommand{\sgn}{\mbox {sgn}}
\newcommand{\var}{\mbox {var}}
\newcommand{\E}{\mbox {E}}
\newcommand{\cov}{\mbox {cov}}
\renewcommand{\Re}{\mbox {Re}}
\renewcommand{\Im}{\mbox {Im}}
\newcommand{\cum}{\mbox {cum}}

\renewcommand{\vec}[1]{{\bf{#1}}}     
\newcommand{\vecsc}[1]{\mbox {\boldmath \scriptsize $#1$}}     
\newcommand{\itvec}[1]{\mbox {\boldmath $#1$}}
\newcommand{\itvecsc}[1]{\mbox {\boldmath $\scriptstyle #1$}}
\newcommand{\gvec}[1]{\mbox{\boldmath $#1$}}

\newcommand{\balpha}{\mbox {\boldmath $\alpha$}}
\newcommand{\bbeta}{\mbox {\boldmath $\beta$}}
\newcommand{\bgamma}{\mbox {\boldmath $\gamma$}}
\newcommand{\bdelta}{\mbox {\boldmath $\delta$}}
\newcommand{\bepsilon}{\mbox {\boldmath $\epsilon$}}
\newcommand{\bvarepsilon}{\mbox {\boldmath $\varepsilon$}}
\newcommand{\bzeta}{\mbox {\boldmath $\zeta$}}
\newcommand{\boldeta}{\mbox {\boldmath $\eta$}}
\newcommand{\btheta}{\mbox {\boldmath $\theta$}}
\newcommand{\bvartheta}{\mbox {\boldmath $\vartheta$}}
\newcommand{\biota}{\mbox {\boldmath $\iota$}}
\newcommand{\blambda}{\mbox {\boldmath $\lambda$}}
\newcommand{\bmu}{\mbox {\boldmath $\mu$}}
\newcommand{\bnu}{\mbox {\boldmath $\nu$}}
\newcommand{\bxi}{\mbox {\boldmath $\xi$}}
\newcommand{\bpi}{\mbox {\boldmath $\pi$}}
\newcommand{\bvarpi}{\mbox {\boldmath $\varpi$}}
\newcommand{\brho}{\mbox {\boldmath $\rho$}}
\newcommand{\bvarrho}{\mbox {\boldmath $\varrho$}}
\newcommand{\bsigma}{\mbox {\boldmath $\sigma$}}
\newcommand{\bvarsigma}{\mbox {\boldmath $\varsigma$}}
\newcommand{\btau}{\mbox {\boldmath $\tau$}}
\newcommand{\bupsilon}{\mbox {\boldmath $\upsilon$}}
\newcommand{\bphi}{\mbox {\boldmath $\phi$}}
\newcommand{\bvarphi}{\mbox {\boldmath $\varphi$}}
\newcommand{\bchi}{\mbox {\boldmath $\chi$}}
\newcommand{\bpsi}{\mbox {\boldmath $\psi$}}
\newcommand{\bomega}{\mbox {\boldmath $\omega$}}

\newcommand{\bolda}{\mbox {\boldmath $a$}}
\newcommand{\bb}{\mbox {\boldmath $b$}}
\newcommand{\bc}{\mbox {\boldmath $c$}}
\newcommand{\bd}{\mbox {\boldmath $d$}}
\newcommand{\bolde}{\mbox {\boldmath $e$}}
\newcommand{\boldf}{\mbox {\boldmath $f$}}
\newcommand{\bg}{\mbox {\boldmath $g$}}
\newcommand{\bh}{\mbox {\boldmath $h$}}
\newcommand{\bp}{\mbox {\boldmath $p$}}
\newcommand{\bq}{\mbox {\boldmath $q$}}
\newcommand{\br}{\mbox {\boldmath $r$}}
\newcommand{\bt}{\mbox {\boldmath $t$}}
\newcommand{\bu}{\mbox {\boldmath $u$}}
\newcommand{\bv}{\mbox {\boldmath $v$}}
\newcommand{\bw}{\mbox {\boldmath $w$}}
\newcommand{\bx}{\mbox {\boldmath $x$}}
\newcommand{\by}{\mbox {\boldmath $y$}}
\newcommand{\bz}{\mbox {\boldmath $z$}}

\newenvironment{Ex}
{\begin{adjustwidth}{0.04\linewidth}{0cm}
\begingroup\small
\vspace{-1.0em}
\raisebox{-.2em}{\rule{\linewidth}{0.3pt}}
\begin{example}
}
{
\end{example}
\vspace{-5mm}
\rule{\linewidth}{0.3pt}
\endgroup
\end{adjustwidth}}

\newcommand{\Hossein}[1]{{\textcolor{blue}{\emph{**Hossein: #1**}}}}
\newcommand{\HosseinC}[1]{{\textcolor{blue}{**Hossein: #1**}}}
\newcommand{\Carlo}[1]{{\textcolor{red}{\emph{**Carlo: #1**}}}}
\newcommand{\Challenge}[1]{{\textcolor{red}{#1}}}
\newcommand{\NEW}[1]{{\textcolor{blue}{#1}}}
\newcommand{\new}[1]{{\textcolor{blue}{#1}}}


\makeatletter
\let\save@mathaccent\mathaccent
\newcommand*\if@single[3]{%
  \setbox0\hbox{${\mathaccent"0362{#1}}^H$}%
  \setbox2\hbox{${\mathaccent"0362{\kern0pt#1}}^H$}%
  \ifdim\ht0=\ht2 #3\else #2\fi
  }
\newcommand*\rel@kern[1]{\kern#1\dimexpr\macc@kerna}
\newcommand*\widebar[1]{\@ifnextchar^{{\wide@bar{#1}{0}}}{\wide@bar{#1}{1}}}
\newcommand*\wide@bar[2]{\if@single{#1}{\wide@bar@{#1}{#2}{1}}{\wide@bar@{#1}{#2}{2}}}
\newcommand*\wide@bar@[3]{%
  \begingroup
  \def\mathaccent##1##2{%
    \let\mathaccent\save@mathaccent
    \if#32 \let\macc@nucleus\first@char \fi
    \setbox\z@\hbox{$\macc@style{\macc@nucleus}_{}$}%
    \setbox\tw@\hbox{$\macc@style{\macc@nucleus}{}_{}$}%
    \dimen@\wd\tw@
    \advance\dimen@-\wd\z@
    \divide\dimen@ 3
    \@tempdima\wd\tw@
    \advance\@tempdima-\scriptspace
    \divide\@tempdima 10
    \advance\dimen@-\@tempdima
    \ifdim\dimen@>\z@ \dimen@0pt\fi
    \rel@kern{0.6}\kern-\dimen@
    \if#31
      \overline{\rel@kern{-0.6}\kern\dimen@\macc@nucleus\rel@kern{0.4}\kern\dimen@}%
      \advance\dimen@0.4\dimexpr\macc@kerna
      \let\final@kern#2%
      \ifdim\dimen@<\z@ \let\final@kern1\fi
      \if\final@kern1 \kern-\dimen@\fi
    \else
      \overline{\rel@kern{-0.6}\kern\dimen@#1}%
    \fi
  }%
  \macc@depth\@ne
  \let\math@bgroup\@empty \let\math@egroup\macc@set@skewchar
  \mathsurround\z@ \frozen@everymath{\mathgroup\macc@group\relax}%
  \macc@set@skewchar\relax
  \let\mathaccentV\macc@nested@a
  \if#31
    \macc@nested@a\relax111{#1}%
  \else
    \def\gobble@till@marker##1\endmarker{}%
    \futurelet\first@char\gobble@till@marker#1\endmarker
    \ifcat\noexpand\first@char A\else
      \def\first@char{}%
    \fi
    \macc@nested@a\relax111{\first@char}%
  \fi
  \endgroup
}
\makeatother

\maketitle

\vspace{-7mm}

\begin{abstract}
In wireless communication networks, interference models are routinely used for tasks such as performance analysis, optimization, and protocol design. These tasks are heavily affected by the accuracy and tractability of the interference models. Yet, quantifying the accuracy of these models remains a major challenge.
In this paper, we propose a new index for assessing the accuracy of any interference model under any network scenario. Specifically, it is based on a new index that quantifies the ability of any interference model in correctly predicting harmful interference events, that is, link outages. We consider specific wireless scenario of both conventional sub-6~GHz and millimeter-wave (mmWave) networks and demonstrate how our index yields insights into the possibility of simplifying the set of dominant interferers, replacing a Nakagami or Rayleigh random fading by an equivalent deterministic channel, and ignoring antenna sidelobes.
Our analysis reveals that in highly directional antenna settings with obstructions, even simple interference models (such as the classical protocol model) are accurate, while with omnidirectional antennas, more sophisticated and complex interference models (such as the classical physical model) are necessary. We further use the proposed index to develop a simple interference model for mmWave networks that can significantly simplify design principles of the important procedures for wireless communication, such as beamforming, interference management, scheduling, and topology control. Our new approach makes it possible to adopt the simplest interference model of adequate accuracy for every wireless network.
\end{abstract}

\begin{IEEEkeywords}
Wireless communications, interference model, performance analysis, millimeter wave networks.
\end{IEEEkeywords}

\section{Introduction}\label{sec: Intro-1}
Due to the shared nature of a wireless media, interference plays a critical role in the design and performance analysis of wireless networks, where the intended signal is combined with other undesired wireless signals transmitted at the same (time, frequency, spatial) channel. The receiver typically decodes the received signal by canceling parts of the interference and treating the rest as noise.
Successful decoding at the receiver depends on the desired signal strength, the ambient noise level accumulated over the operating bandwidth, and the interference level. Signal-to-interference-plus-noise ratio (SINR) is a common metric to evaluate the outage probability (or the probability of successful decoding) of a transmission. However, performance analysis using the SINR expression is complex as it depends on the transmission strategies (transmission power, antenna pattern, and medium access control (MAC) protocol), often unknown or hard to estimate random channel attenuation, receiver design, and the (often partially unknown) network topology.
Due to this overwhelming complexity, the design and analysis of wireless networks based on the actual SINR expression, while being accurate, is very challenging. This difficulty is further exacerbated in millimeter-wave (mmWave) networks, where penetration loss, first-order reflection, and antenna pattern introduce further elements of randomness~\cite{Akdeniz2014MillimeterWave,shokri2015mmWavecellular,di2014stochastic}.
This motivates developing different techniques to mathematically model (abstract) various components of the SINR, e.g., the transmission strategy, wireless channel, and network topology.

\subsection{Related Works and Motivations}\label{sec: motivations}
Define an \emph{interference model} as a set of deterministic or stochastic functions that model various components of the SINR expression. There have been many attempts in the literature to design interference models (equivalently, to approximate the SINR expression) that accurately capture the effect of interference, while being tractable for the mathematical analysis. These interference models largely try to answer the following questions under various network settings:
\begin{itemize}
\item[{Q1.}] How can we model the set of interferer whose contributions in the aggregated interference term are dominant?
\item[{Q2.}] How can we simplify the transmission/reception and propagation models to enhance tractability of the interference model with marginal loss in its accuracy?
\end{itemize}

Answering Q1 demands a careful balance between the accuracy and the simplicity of the interference model. Considering the effects of more interferers in the SINR model generally increases the accuracy but also the complexity. In this regard, the simplest model is the \emph{primary interference model}~\cite{ephremides1987design}, wherein an outage event occurs only if two communication links share a common endpoint. In other words, the only interference component in this model is self-interference that leads to a half-duplex operating mode. \emph{Interference range model} (IRM) is an attempt to improve the accuracy of the primary interference model~\cite{iyer2009right}, where an outage event occurs if the closest interferer is located no farther than a certain distance of the receiver, called the interference range. By setting this distance to 0, the IRM can be reduced to the primary interference model. A modified version of IRM is the \emph{protocol model} (PRM), formalized by the seminal work of Gupta and Kumar~\cite{gupta2000capacity}. The only modification is that the interference range, instead of being a constant value as in the IRM, depends on the received power from the intended transmitter and a minimum SINR threshold for successful decoding. Although the IRM and PRM are very simple, they fail to capture the effect of interference aggregation (i.e., the sum of the interference power from multiple interferers). It might be that, while there is no interferer inside the interference range, the aggregated interference from several transmitters outside the interference range downs the perceived SINR below the threshold. Thus, these models are generally considered to be overly simplistic. Nonetheless, due to their mathematical tractability, the IRM (including the primary interference model) and PRM are extensively adopted for the performance analysis and for the system design; e.g., transport capacity~\cite{gupta2000capacity,liu2003capacity,kyasanur2009capacity}, delay~\cite{gamal2004throughput,el2006optimal}, fairness~\cite{nandagopal2000achieving}, throughput~\cite{xu2002effective,Singh2011Interference,Yu2016Distributed}, topology control~\cite{marina2010topology,Stahlbuhk2016Topology}, routing~\cite{jain2005impact}, and backoff design~\cite{celik2010mac}.

To alleviate the aforementioned problem of IRM and PRM, the \emph{interference ball model} (IBM) considers the aggregated impacts of near-field interferers, located no farther than a certain distance. The price is higher complexity of the IBM compared to the IRM and PRM. Nonetheless, the IBM has been extensively adopted in the performance evaluation of wireless networks~\cite{weber2005transmission,weber2007transmission,di2014stochastic,le2010longest}. The topological interference model (TIM)~\cite{jafar2014topological} is a natural extension of the IBM that considers the aggregated impact of all the transmitters whose individual interference level at the receiver side is not below a certain threshold. In other words, this model neglects weak links based on the ``topological'' knowledge. The TIM is adopted for capacity and degree-of-freedom analysis~\cite{jafar2014topological,yi2014topological}. The most accurate and complex answer to Q1 is the \emph{physical model} (PhyM)~\cite{gupta2000capacity}, which considers the aggregated interference of all transmitters in the entire network.\footnote{Under very special network settings (e.g., homogenous Poisson field of interferers exhibiting Rayleigh fading channel), the PhyM may be mathematically more tractable than both PRM and IBM~\cite{Haenggi2013Stochastic}; however, the PRM and IBM are yet more desirable models for protocol design and for network optimization~\cite{le2010longest}.} The PhyM, also known as the SINR model, is adopted mostly at the physical layer; e.g., beamforming design~\cite{schubert2004solution,dahrouj2010coordinated,Moghadam2017PilotJSAC}, capacity evaluation~\cite{gupta2000capacity,sharif2005capacity,baccelli2009stochastic}, power control~\cite{rashid1998joint,chandrasekhar2009power}, coverage analysis~\cite{di2014stochastic}, energy efficiency characterization~\cite{ngo2013energy}, and spectrum sharing~\cite{shokri2016Spectrum}.

The answer to Q2 depends heavily on the transmission and reception strategies and propagation environment. For instance, approximating the random wireless channel gain with its first moment (average) is a common technique to simplify the SINR expression and to design MAC and networking layers~\cite{cardieri2010modeling,weber2007transmission,badia2008general,chen2006cross,Singh2011Interference,Singh2009Blockage,Stahlbuhk2016Topology}. Reference~\cite{Gupta2016Sharing} replaced a Nakagami fading channel by a Rayleigh one for mathematical tractability and numerically concluded from its Fig.~5 that such approximation preserves the main properties of the rate coverage performance. Yet, the impact of these mathematical approximations on the accuracy of the performance analysis is not well understood. Recently,~\cite{Dams2015Scheduling} considers the impact of such approximation on the scheduling. In particular, the authors show that, if we design scheduling for $n$ transmitters based on a proper non-fading channel model (deterministic approximation of the random channel gain), the network throughput will be within $\mathcal{O}(\log n)$ of that of the optimal scheduler, designed based on the actual random channel gains. This result, however, is limited to the Rayleigh fading model. As another example, for mmWave communications with many antenna elements,~\cite{Singh2011Interference} and~\cite{Shokri2015Transitional} assume no emissions from the antenna sidelobe, which affects the SINR distribution. This assumption is relaxed in~\cite{di2014stochastic}, where the antenna sidelobe is modeled by a small constant value, adding further complexity into the interference model. As a result, the final derivations, while being more accurate, are less tractable and provide less insights. However, without having a mathematical framework that allows assessing the impact of neglecting antenna sidelobes, it is not clear which approach better balances the simplicity-accuracy tradeoff of mathematical analysis.

The proper choice of interference model depends on many parameters such as the receiver design, antenna directionality, network topology, channel model, and the choice of medium access protocol~\cite{badia2008general,iyer2009right,cardieri2010modeling}. To the best of our knowledge, there has been no systematic method to \textit{analyze the accuracy of various interference models, choose the proper interference model, and quantify the amount of error due to adopting other interference models for a given network scenario}. The accuracy of different interference models has been mostly evaluated qualitatively, without fully understanding the mutual impacts of different parameters of the physical, medium access, and network layers. This qualitative analysis, however, is often overly simplistic, and may result in the use of interference models that are only marginally more accurate, yet significantly more complex than needed. As we will show throughout this paper, in certain settings of relevant practical interest, even the simplest interference models are sufficiently accurate and can be used to provide significant insights into the network performance and to enable efficient protocol design.

\subsection{Contributions}
In this paper, we substantially extend the preliminary version of this study~\cite{Shokri2016OntheAccuracy} and propose a new framework to assess the distance of two arbitrary SINR distributions. We use this framework to develop an interference model similarity index that takes on real values between 0 and 1, where higher values correspond to higher similarity. This index builds a universal method to assess the accuracy of any interference model under any network scenario. In other words, instead of introducing a new interference model or a new approach to analyze SINR distribution, we propose a novel framework to investigate the accuracy of the existing interference models. Therefore, our study is complementary to the rich literature of interference analysis.

To exemplify the abilities of the proposed index, we mathematically evaluate it for the PRM and IBM under three scenarios: (\textit{i}) Rayleigh fading channel and omnidirectional communications (a typical sub-6~GHz system); and (\textit{ii}) Rayleigh fading channel and directional communications; and (\textit{iii}) deterministic wireless channel, directional communications, and existence of impenetrable obstacles in the environment (a typical mmWave system). Although the applications of the proposed index is general and goes beyond the examples provided in this paper, we use these examples to illustrate fundamental properties of this index and also to provide insights on the mutual effects of various network parameters on the accuracy of the interference model, thus commenting on the proper model for a given network scenario.

In the first example scenario, served as a baseline, we derive a closed-form expression for the accuracy index. We show that the accuracy of the IBM monotonically increases with the interference range, at the expense of an increased complexity. In contrast, we show that there is no such monotonic improvement in the accuracy of PRM. Thereby, we find the optimal interference range that maximizes the accuracy of the PRM.

In the second example scenario, we show that both the PRM and IBM are significantly more accurate with directional antennas. Further, in the third example scenario, we show significant accuracy improvement of both PRM and IBM due to deterministic channel, directionality, and also blockage. As these conditions hold in mmWave networks, we show that the PRM can be used in the analysis of mmWave networks to significantly improve the mathematical tractability of the problem, with a negligible loss in the analysis accuracy. We further use this index to observe marginal impacts of the first-order reflection and sidelobe transmissions on the accuracy of the interference model, which inspire us to propose a tractable and accurate interference model for mmWave networks.

Furthermore, we use the proposed framework to investigate the feasibility of modeling a random fading channel with a deterministic channel. We show that if the spatial distribution of the transmitters follow a Poisson point process on the plane and if the path-loss exponent is 2, then the average of the fading random variable\footnote{Rigorously speaking, the fading should be an absolutely continuous random variable, which holds for almost all wireless channels.} is among the best constant approximations of the random fading channel to analyze any ergodic function of the SINR (e.g., transport capacity, throughput, and delay).

Throughout the paper, we show how the proposed index can increase our understanding of the mutual interactions among the accuracy of the performance evaluation and various network parameters and modeling techniques. We also signify how we can rigorously develop simple interference models of adequate accuracy to simplify design principles of the main functions of wireless communications such as beamforming, interference management, scheduling, and topology control.

The rest of this paper is organized as follows. In Section~\ref{sec: system-model}, we introduce our interference model similarity index, and investigate it under various network scenarios in Sections~\ref{sec: Rayleigh-noBlockage}--\ref{sec: FurtherDiscussions}. Future works are presented in Section~\ref{sec: future_directions}, and the paper is concluded in Section~\ref{sec: ConcludingRemarks}.

\section{Interference Model Similarity Index}\label{sec: system-model}

\subsection{Interference Model}
We define a link as the pair of a transmitter and its intended receiver, where transmitter (receiver) $i$ refers to the transmitter (receiver) of link $i$. Without loss of generality and for brevity, we assume that there is no interference cancellation, so all unintended transmitters act as potential interferers to any receiver. Consider a reference receiver and label its intended transmitter by subscript 0. Denote by $\mathcal{I}$ the set of its interferers (all active transmitters excluding the intended transmitter), by $p_i$ the transmission power of transmitter $i$, by $\sigma$ the power of white Gaussian noise, by $d_{i}$ the distance between transmitter $i$ and the reference receiver, and by $g_{i}^{\mathrm{Ch}}$ the channel gain between transmitter $i$ and the reference receiver.
We denote by $g_{i}^{\mathrm{Tx}}$ the antenna gain at transmitter $i$ toward the reference receiver, and by $g_{i}^{\mathrm{Rx}}$ the antenna gain at the reference receiver toward transmitter $i$. Then, the SINR at the reference receiver is
\begin{equation*}
\gamma = \frac{p_0 g_{0}^{\mathrm{Tx}} g_{0}^{\mathrm{Ch}} g_{0}^{\mathrm{Rx}}}{\sum\limits_{k \in \mathcal{I}} p_k g_{k}^{\mathrm{Tx}} g_{k}^{\mathrm{Ch}} g_{k}^{\mathrm{Rx}} + \sigma} \:.
\end{equation*}
The SINR depends on the transmission powers, antenna patterns, set of active transmitters, channel model, and network topology. Let $\beta>0$ denote the SINR threshold corresponding to a certain target bit error rate. An outage on the reference link occurs when $\gamma < \beta$. Different interference models attempt to approximate the outage probability by ignoring certain components of the interference (see questions Q1 and Q2 in Section~\ref{sec: motivations}). In particular, the IRM, PRM, IBM, TIM, and PhyM characterize the set of interferers $\mathcal{I}$. Neglecting various components of the channel model translates into different distributions for $g_{i}^{\mathrm{Ch}}$.
Power allocation affects $p_i$, and various scheduling protocols further affect $\mathcal{I}$.

\subsection{Formal Definition of the Similarity Index}
Consider reference interference model $\mathrm{y}$ under a given set of parameters/functions describing the wireless network. Define $\gamma^{\mathrm{y}}$ as the SINR of a reference receiver under this model. We define a binary hypothesis test, where hypotheses $H_0$ and $H_1$ denote the absence and presence of outage under reference model $\mathrm{y}$, respectively. That is,
\begin{equation}\label{eq: binary-hypothesis-test}
\begin{cases}
  H_0, & \mbox{if}~ \gamma^{\mathrm{y}} \geq \beta \:,\\
  H_1, & \mbox{if}~ \gamma^{\mathrm{y}} < \beta \:.
\end{cases}
\end{equation}
We consider a test interference model $\mathrm{x}$ under any set of parameters/functions describing our wireless network, which are not necessarily equal to those of the reference model $\mathrm{y}$. These differences result in possible deviation of the SINR of the reference receiver under $\mathrm{x}$, denoted by $\gamma^{\mathrm{x}}$, from $\gamma^{\mathrm{y}}$. From the outage point of view, irrespective of the differences between individual parameters/functions of $\mathrm{x}$  and $\mathrm{y}$, we say model $\mathrm{x}$ is similar to model $\mathrm{y}$ if it gives exactly the same outage result as $\mathrm{y}$. Assume interference model $\mathrm{x}$ is a detector of outage events under $\mathrm{y}$. To evaluate the performance of this detector compared to reference model $\mathrm{y}$, we can use the notions of false alarm and miss-detection. A false alarm corresponds to the event that $\mathrm{x}$ predicts outage under hypothesis $H_0$ (i.e., $\mathrm{y}$ declares no harmful interference); whereas a miss-detection corresponds to the event that $\mathrm{x}$ fails to predict outage under hypothesis $H_1$. Now, the performance of any interference model $\mathrm{x}$ can be evaluated using the false alarm and miss-detection probabilities, namely $p_{\mathrm{fa}}^{\mathrm{x} \mid \mathrm{y}}$ and $p_{\mathrm{md}}^{\mathrm{x} \mid \mathrm{y}}$. Formally,
\begin{align}\label{eq: miss-detection-prob}
p_{\mathrm{fa}}^{\mathrm{x} \mid \mathrm{y}} &= \Pr \left[\gamma^{\mathrm{x}} < \beta \mid \gamma^{\mathrm{y}} \geq \beta \right] \:,
\quad p_{\mathrm{md}}^{\mathrm{x} \mid \mathrm{y}} = \Pr \left[\gamma^{\mathrm{x}} \geq \beta \mid \gamma^{\mathrm{y}} < \beta \right] \:.
\end{align}
The false alarm and miss-detection probabilities quantify the similarity of any interference model $\mathrm{x}$ in detecting outage events compared to any reference model $\mathrm{y}$. Next, we define our index to be a convex combination of these probabilities.

\begin{defin}[Interference Model Similarity Index]\label{def: IMS-index}
For any constant $0 \leq \xi \leq 1$, any SINR threshold $\beta$, any test interference model $\mathrm{x}$, and any reference interference model $\mathrm{y}$, we define similarity of $\mathrm{x}$ to $\mathrm{y}$ at $\beta$ as
\begin{align}\label{eq: Definition-IMSindex}
S_{\beta,\xi}\left(\mathrm{x}\|\mathrm{y} \right)  = \xi \left( 1 - p_{\mathrm{fa}}^{\mathrm{x} \mid \mathrm{y}}\right) + \left( 1 - \xi \right) \left( 1 - p_{\mathrm{md}}^{\mathrm{x} \mid \mathrm{y}}\right) = 1 - \xi \, p_{\mathrm{fa}}^{\mathrm{x} \mid \mathrm{y}} - \left( 1 - \xi \right) p_{\mathrm{md}}^{\mathrm{x} \mid \mathrm{y}} \:,
\end{align}
where $p_{\mathrm{fa}}^{\mathrm{x} \mid \mathrm{y}}$ and $p_{\mathrm{md}}^{\mathrm{x} \mid \mathrm{y}}$ are given in~\eqref{eq: miss-detection-prob}. Notice that random variables $\gamma^{\mathrm{x}}$ and $\gamma^{\mathrm{y}}$ must have a common support.
\end{defin}

$S_{\beta,\xi}\left(\mathrm{x}\|\mathrm{y} \right)$ is a unit-less  quantity ranging within $[0,1]$, where higher values represents higher similarity between $\mathrm{x}$ and $\mathrm{y}$ in capturing outage events at SINR threshold $\beta$. Setting $\xi = \Pr \left[ \gamma^{\mathrm{y}} \geq \beta \right]$, $\xi p_{\mathrm{fa}}^{\mathrm{x} \mid \mathrm{y}} + \left( 1 - \xi \right) p_{\mathrm{md}}^{\mathrm{x} \mid \mathrm{y}}$ is the average error in detecting the outage events; therefore, $S_{\beta,\Pr \left[\gamma^{\mathrm{y}} \geq \beta \right]}\left(\mathrm{x}\|\mathrm{y} \right)$ shows the probability that interference model $\mathrm{x}$ has similar decision as reference interference model $\mathrm{y}$ in detecting the outage events.
\begin{remark}[Accuracy of an Interference Model]\label{result: accuracy-index}
Let reference model $\mathrm{y}$ perfectly capture the outage events in reality, namely the model $\mathrm{y}$ does not make any approximation/simplification. The accuracy of any interference model $\mathrm{x}$ is then $S_{\beta,\xi}\left(\mathrm{x}\|\mathrm{y} \right)$, and we call it the \emph{accuracy index} throughout the paper.
\end{remark}
The proposed index is a universal metric that can be used to quantify the accuracy of any interference models, proposed in the literature, as we exemplify in the following sections.

\subsection{Comparison to the Existing Statistical Distance Measures}
Interference model similarity index, formulated in~\eqref{eq: Definition-IMSindex}, is measuring the distance\footnote{Rigorously speaking, our similarity index is not a distance measure, as it does not satisfy the subadditivity property. Moreover, we are measuring the similarity, which could be in general a decreasing function of the distance.} of the PDF of $\gamma^{\mathrm{x}}$ compared to that of $\gamma^{\mathrm{y}}$. Let $f_{X}$ denote PDF of random variable $X$.
In the following, we highlight three main advantages of using our index with respect to the existing standard distance measures, such as the Bhattacharyya distance and the Kullback-Leibler (KL) divergence~\cite{Kailath1967Bhattacharyya}.

First, the existing standard distance measures mostly map the distance between $f_{\gamma^{\mathrm{x}}}$ and $f_{\gamma^{\mathrm{y}}}$ in their entire support to only one real value. It might be that two distribution are very similar in the meaningful ranges of the SINR values (0--10~dB), but very different outside this range. Still, the classical statistical distance measures may result in a high distance between two distributions, as they compare $f_{\gamma^{\mathrm{x}}}$ to $f_{\gamma^{\mathrm{y}}}$ in the entire SINR range. This is indeed a misleading result that may mistakenly avoid the use of the simplified interference model $\mathrm{x}$ in practice.
However, our similarity index allows us to investigate whether or not $\mathrm{x}$ is accurate at any given SINR threshold.

Second, both the Bhattacharyya distance and the KL divergence may fail in a comparative analysis. In particular, $f_{\gamma^{\mathrm{y}}}$ might be more similar to $f_{\gamma^{\mathrm{x}}}$ than $f_{\gamma^{\mathrm{z}}}$ with point-wise comparison, but the  Bhattacharyya distance and the KL divergence of $f_{\gamma^{\mathrm{x}}}$ from $f_{\gamma^{\mathrm{y}}}$ become higher than that of $f_{\gamma^{\mathrm{z}}}$ from $f_{\gamma^{\mathrm{y}}}$, as shown in the following toy example.

\begin{example}
Consider discrete random variables $\mathrm{X}$, $\mathrm{Y}$, and $\mathrm{Z}$ with common support of $[1,2,3]$ with probability mass functions

{\begin{table}[!h]
\centering
{
\renewcommand{\tabcolsep}{5pt}
\renewcommand{\arraystretch}{1}
\begin{tabular}{|c|c|c|c|}
  \hline
  t & $f_{X}(t)$ & $f_{Y}(t)$ & $f_{Z}(t)$ \\ \hline
  1 & 0.05 & 0.1 & 0.25 \\  \hline
  2 & 0.25 & 0.45 & 0.2 \\  \hline
  3 & 0.7 & 0.45 & 0.55 \\
  \hline
\end{tabular}
}
\end{table}}
\vspace{5mm}

Then, we have the following metrics ($f_{X}$ is the reference in the KL divergence):
{\begin{table}[!h]
\centering
{
\renewcommand{\tabcolsep}{5pt}
\renewcommand{\arraystretch}{1}
\begin{tabular}{|l|c|c|}
  \hline
  Distributions & $f_{X}$,$f_{Y}$ & $f_{X}$,$f_{Z}$ \\ \hline
  Euclidean distance & 0.324 & 0.255 \\  \hline
  Bhattacharyya distance & 0.033 & 0.045 \\  \hline
  KL divergence & 0.059 & 0.098 \\  \hline
\end{tabular}
}
\vspace{-7mm}
\end{table}
}
\linebreak
In this example, neither the Bhattacharyya distance nor the KL divergence can identify higher point-wise similarity of $Z$ to $X$ than $Y$ to $X$.
\end{example}

Last, but not least, unlike the existing statistical distance metrics that are not necessarily intended for communication systems, our similarity index is developed for these systems so that it has a physical meaning and can provide practical insights. Specifically, setting $\xi = \Pr \left[ \gamma^{\mathrm{y}} \geq \beta \right]$, our index $S_{\beta,\xi}\left(\mathrm{x}\|\mathrm{y} \right)$ evaluates the probability of correct decision of outage events under interference model $\mathrm{x}$.

Note that other distance metrics may still be useful to evaluate the accuracy of an interference model, and they may also have some relationship to our proposed index; see the following remark as an example.
\begin{remark}[Relationship to the Bhattacharyya Coefficient]\label{remark: Bhattacharyya-distance}
Let $\xi = \Pr \left[\gamma^{\mathrm{y}} \geq \beta \right]$. By noting that $S_{\beta,\Pr \left[\gamma^{\mathrm{y}} \geq \beta \right]}\left(\mathrm{x}\|\mathrm{y} \right)$ is the probability of having no hypothesis detection error and following~\cite[Equation~(48)]{Kailath1967Bhattacharyya}, we get
\begin{align}\label{eq: ComparisonWithBhattacharyya}
\frac{3}{2} - \xi - \rho \sqrt{\xi \left(1-\xi\right)} \leq &~ S_{\beta,\xi}\left(\mathrm{x}\|\mathrm{y} \right)\leq 1 - \xi + \sqrt{\frac{1}{4} - \xi \left(1-\xi\right) \rho^2} \:,
\end{align}
where $\rho = \int{f_{\gamma^{\mathrm{x}}}(t)f_{\gamma^{\mathrm{y}}}(t)\mathrm{d}t}$ is the Bhattacharyya coefficient.
\end{remark}

\subsection{Applications of the Interference Model Similarity Index}\label{sec: set-of-interf}
In the following, we provide two class of illustrative examples where our index can be used either to simplify the mathematical analysis or to justify the existing interference models. Use cases of our index, however, goes beyond these examples.
\subsubsection{Simplifying the Set of Interferers}
This is one of the first steps in choosing an interference model for performance analysis, protocol design, and network optimization. With omnidirectional transmission/reception and without interference cancelation, an outage occurs under
\begin{itemize}
  \item PRM: if there is an active transmitter no farther than an interference range $r_{\mathrm{PRM}} = (1+\Delta)d_{0}$, where $\Delta$ is a constant real positive value~\cite{gupta2000capacity};
  \item IBM: if its SINR due to all active transmitters located no farther than an interference range $r_{\mathrm{IBM}}$ is less than $\beta$~\cite{le2010longest};
  \item TIM: if its SINR due to all active transmitters with strong links (with individual channel gains higher than $\varepsilon$) toward receiver $i$ is less than $\beta$~\cite{jafar2014topological}; and
  \item PhyM: if its SINR due to all active transmitters is less than $\beta$~\cite{gupta2000capacity}.
\end{itemize}
To present a unified view, we associate three random variables $a_{k}^{\mathrm{PRM}}$, $a_{k}^{\mathrm{IBM}}$, and $a_{k}^{\mathrm{TIM}}$ to the link between each transmitter $k \in \mathcal{I}$ and the typical receiver. $a_{k}^{\mathrm{PRM}}$ is set ${+\infty}$ if $d_{k} \leq (1+\Delta)d_{0}$, and otherwise 0. $a_{k}^{\mathrm{IBM}}$ is set $1$ if $d_{k} \leq r_{\mathrm{IBM}}$, and otherwise 0. Finally, $a_{k}^{\mathrm{TIM}}$ is set $1$ if $g_{k}^{\mathrm{Ch}} > \varepsilon $, and otherwise 0. We define a virtual channel gain for those interference models as
\begin{equation}\label{eq: general-model}
g_{k}^{\mathrm{x}} = a_{k}^{\mathrm{x}} g_{k}^{\mathrm{Ch}}\:, \qquad  \mbox{for interference model}~\mathrm{x} \:,
\end{equation}
where $\mathrm{x}$ is a label denoting PRM, IBM, TIM, or PhyM, and $a_{i}^{\mathrm{PhyM}} \triangleq 1$. Despite the virtual channel gain, all other parameters of interference models $\mathrm{x}$ and $\mathrm{y}$ are identical. The SINR at the typical receiver under interference model $\mathrm{x}$ is given by
\begin{equation}\label{eq: protocol-model}
\gamma^{\mathrm{x}} = \frac{p_0 g_{0}^{\mathrm{Tx}} g_{0}^{\mathrm{Ch}} g_{0}^{\mathrm{Rx}}}{\sum\limits_{k \in \mathcal{I}} p_k g_{k}^{\mathrm{Tx}} g_{k}^{\mathrm{x}} g_{k}^{\mathrm{Rx}} + \sigma}  \:.
\end{equation}
The design of many key functions of a wireless network such as scheduling~\cite{modiano2006maximizing} or power allocation~\cite{rashid1998joint} need an estimate of~\eqref{eq: protocol-model}. To this end,
a receiver may need to coordinate with a set of interferers to estimate their individual instantaneous contributions to the SINR expression, namely $p_k g_{k}^{\mathrm{Tx}} g_{k}^{\mathrm{x}} g_{k}^{\mathrm{Rx}}$ for all $k\in \mathcal{I}$. The PhyM may imply that every receiver should coordinate with all the interferers in the entire network (global information) whose cost, complexity, and delay may be unaffordable in many networking scenarios. Using IBM implies that each nodes should coordinate with all transmitters within a certain radius (local information), and the PRM necessitates coordination only with the closest unintended transmitter, which are appealing from energy and protocol overhead perspectives. Our proposed index gives a quantitative insight on the accuracy of various interference models, used for protocol development and for network optimization, and allows the use of the right interference model for a given channel model and network scenario.


\subsubsection{Simplifying the Channel Model}
Our accuracy index can be used to adopt tractable channel models ($g_{k}^{\mathrm{Ch}}$ for every transmitter $k$) of adequate accuracy. This is specially important for mmWave networks, where LoS and non-LoS conditions have different channel models, non-LoS (blockage) probability follows a rather complicated function, the LoS channel may follow a Nakagami fading in general, and realistic antenna patterns might be a complicated non-linear function. Various researches tried to simplify those complications without rigorous analysis on the validity of such simplifications. For instance,~\cite{Singh2011Interference} assumed impenetrable obstacles (so communication only in the LoS conditions) and neglected antenna sidelobe,~\cite{di2014stochastic} approximated the non-LoS stochastic function by a deterministic LoS ball in which there is no obstacle within a certain range of the receiver and there is no LoS links outside the circle, and~\cite{Gupta2016Sharing} replaced the Nakagami fading channel by a Rayleigh fading that facilitates mathematical analysis. Due to lack of a systematic approach to simplify the channel model, the understanding of the cross-layer dynamics between MAC and physical layers of most of the existing standards is a largely open problem, and the existing frameworks such as the one in~\cite{PG2014modeling} are not usually mathematically tractable.

In the following, we illustrate the utility of our index for four example scenarios. Although our index poses no limitation to these example scenarios, we may simplify some parameters of the system model to avoid unnecessary complications.
In the first three examples, we focus on simplifying the set of interferers for various network settings and derive closed-form expressions for the accuracy index to highlight its fundamental properties. In the last example scenario, we use our index to numerically assess the accuracy of various approaches in simplifying the channel model.

For the rest of this paper, without loss of generality, we assume $\xi = \Pr \left[ \gamma^{\mathrm{y}} \geq \beta \right]$, so $S_{\beta,\xi}\left(\mathrm{x}\|\mathrm{y} \right)$ evaluates the probability of correct decision under interference model $\mathrm{x}$.

\section{Example Scenario 1: Rayleigh Fading Channel with Omnidirectional Communications}\label{sec: Rayleigh-noBlockage}
Consider a wireless network with Rayleigh fading channel and omnidirectional transmission/reception. Assume that the PhyM can perfectly capture the outage events. In this section, we evaluate the accuracy of IBM, PRM, and TIM (see Section~\ref{sec: set-of-interf} where we recalled the definition of these prominent models) for such scenario.

We consider a reference receiver (called the typical receiver) at the origin of the Polar coordinate, and its intended transmitter having geometrical/spatial length $d_{0}$. We consider a homogeneous Poisson network of interferers (unintended transmitters) on the plane with intensity $\lambda_t$. We assume that all the transmitters are active with transmission power $p$ (no power control), and that there is no interference cancellation, which are natural assumptions in personal and local area networks. With omnidirectional transmission and reception, there is no antenna gains, so $g_{k}^{\mathrm{Tx}} = g_{k}^{\mathrm{Rx}} = 1$, $k \in \mathcal{I} \cup \{0\}$. Note that, under these assumptions, the PhyM is more tractable for coverage and rate analyses than other models (PRM, IBM, and TIM)~\cite{Haenggi2013Stochastic}; however, we still use this example to derive closed-form expression for the new accuracy index and thereby illustrate its fundamental properties that hold in general. Nonetheless, even in this network setting, the PRM and IBM are more appealing than PhyM for protocol design and for network optimization~\cite{le2010longest}.

We define by $\mathcal{B}(\theta,r_{\mathrm{in}},r_{\mathrm{out}})$ a geometrical annulus sector with angle $\theta$, inner radius $r_{\mathrm{in}}$, and outer radius $r_{\mathrm{out}}$, centered at the location of the typical receiver (origin of the Polar coordinate). To model a wireless channel, we consider a constant attenuation $c$ at reference distance 1~m, a distance-dependent attenuation with exponent $\alpha$, and a Rayleigh fading component $h$. To avoid the physically unreasonable singularity that arises at the origin under power law attenuation, we change the path loss index to $\alpha \mathds{1}_{\overline{\mathcal{B}}(2\pi,0,a)}$, where $\mathds{1}_{\cdot}$ is the indicator function assuming value 1 over set $\cdot$ and zero otherwise. This modified power law model implies that the signal of all transmitters located outside a disk with radius $a$ will be attenuated by traditional power law method; however, the transmitters inside this disk will observe no channel attenuation.  Therefore, the channel gain between transmitter $i$ at radial distance $d_i$ and the typical receiver is $g_{i}^{\mathrm{Ch}} = c h_i d_{i}^{-\alpha \mathds{1}_{\overline{\mathcal{B}}(2\pi,0,a)}}$. To avoid unnecessary complications while illustrating the utility of our index, we eliminate the shadow fading from our channel model.

We are now ready to illustrate the utility of our proposed index using the SINR expression~\eqref{eq: protocol-model}.

\subsection{Accuracy of the Interference Ball Model}\label{subsec: example-1-IBM}
For mathematical tractability, we assume that $r_{\mathrm{IBM}} \geq a$ and $d_0 \geq a$, and the extension to the general case is straightforward.
The false alarm probability can be reformulated as
\begin{align}\label{eq: Pfa-IBM-1}
p_{\mathrm{fa}}^{{\mathrm{IBM}} \mid {\mathrm{PhyM}}} = \Pr \left[\gamma^{\mathrm{IBM}} < \beta \mid \gamma^{\mathrm{PhyM}} \geq \beta \right]  = \frac{\Pr \left[\gamma^{\mathrm{IBM}} < \beta \right] \Pr \left[\gamma^{\mathrm{PhyM}} \geq \beta \mid \gamma^{\mathrm{IBM}} < \beta \right]}{1 - \Pr \left[\gamma^{\mathrm{PhyM}} < \beta \right]} \:.
\end{align}
Although the PhyM considers the impacts of all the interferers in the entire network, the IBM considers only the effects of the near-field ones. Consequently, $\gamma^{\mathrm{PhyM}} \leq \gamma^{\mathrm{IBM}}$, and thus $\Pr \left[\gamma^{\mathrm{PhyM}} \geq \beta \mid \gamma^{\mathrm{IBM}} < \beta \right] = 0$ in the nominator of~\eqref{eq: Pfa-IBM-1}. This results in ${p_{\mathrm{fa}}^{{\mathrm{IBM}} \mid {\mathrm{PhyM}}} = 0}$.

For the miss-detection probability, we have
\begin{align}\label{eq: miss-detection-uWave-Rayleigh-1}
p_{\mathrm{md}}^{{\mathrm{IBM}} \mid {\mathrm{PhyM}}}  &= \Pr \left[\gamma^{\mathrm{IBM}} \geq \beta \mid \gamma^{\mathrm{PhyM}} < \beta \right] = 1 - \Pr \left[\gamma^{\mathrm{IBM}} < \beta \mid \gamma^{\mathrm{PhyM}} < \beta \right] \nonumber \\
& = 1 - \frac{\Pr \left[\gamma^{\mathrm{IBM}} < \beta \right] \Pr \left[\gamma^{\mathrm{PhyM}} < \beta \mid \gamma^{\mathrm{IBM}} < \beta \right]}{\Pr \left[\gamma^{\mathrm{PhyM}} < \beta \right]} = 1 - \frac{\Pr \left[\gamma^{\mathrm{IBM}} < \beta \right]}{\Pr \left[\gamma^{\mathrm{PhyM}} < \beta \right]} \:,
\end{align}
where the last equality is from $\gamma^{\mathrm{PhyM}} \leq \gamma^{\mathrm{IBM}}$. In Appendix~A, we have derived
\begin{align}\label{eq: IBM1-1}
\hspace{-3mm}\Pr\left[\gamma^{\mathrm{IBM}} < \beta \right]  & = 1 - \exp\Vast\{ \frac{- \sigma \beta d_{0}^{\alpha}}{pc} -\pi \lambda_t  \, \mathbf{E}_h \vast[ a^2 \left( 1 - e^{- \beta d_{0}^{\alpha} h} \right) + r_{\mathrm{IBM}}^2 \left( 1 - e^{- \beta d_{0}^{\alpha} h r_{\mathrm{IBM}}^{-\alpha}} \right) -
\vast. \Vast. \nonumber \\[-5mm]
& \hspace{-25mm} \Vast. \vast.  a^2 \left( 1 - e^{- \beta d_{0}^{\alpha} h a^{-\alpha}} \right) + \left( \beta d_{0}^{\alpha} h \right)^{2/\alpha} \Gamma\left( 1 - \frac{2}{\alpha}, \beta d_{0}^{\alpha} h r_{\mathrm{IBM}}^{-\alpha} \right)
- \left( \beta d_{0}^{\alpha} h \right)^{2/\alpha} \Gamma\left( 1 - \frac{2}{\alpha}, \beta d_{0}^{\alpha} h a^{-\alpha} \right)  \hspace{-1.5mm} \vast] \hspace{-1.5mm} \Vast\}  \:,
\end{align}
and
\begin{align}\label{eq: PhyM1-1}
\Pr \left[\gamma^{\mathrm{PhyM}} < \beta \right] &= 1 - \exp\Vast\{ - \frac{\sigma \beta d_{0}^{\alpha}}{pc} - \pi \lambda_t  \, \mathbf{E}_h \vast[ a^2 \left( 1 - e^{- \beta d_{0}^{\alpha} h} \right) - a^2 \left( 1 - e^{- \beta d_{0}^{\alpha} h a^{-\alpha}} \right) \vast. \Vast. \nonumber \\[-5mm]
& \hspace{10mm} \Vast. \vast. + \left( \beta d_{0}^{\alpha} h \right)^{2/\alpha} \Gamma\left( 1 - \frac{2}{\alpha} \right)
- \left( \beta d_{0}^{\alpha} h \right)^{2/\alpha} \Gamma\left( 1 - \frac{2}{\alpha}, \beta d_{0}^{\alpha} h a^{-\alpha} \right) \vast] \Vast\}  \:,
\end{align}
where $\Gamma\left(\cdot, \cdot \right)$ is the incomplete Gamma function, $\Gamma\left(\cdot \right)$ is the Gamma function, $\mathbf{E}_{h}$ denotes expectation over random variable $h$, and the probability density function of $h$ is $f_h(x) = e^{-x}$. Substituting~\eqref{eq: IBM1-1} and~\eqref{eq: PhyM1-1} into~\eqref{eq: miss-detection-uWave-Rayleigh-1}, the miss-detection probability can be found. Also, from~\eqref{eq: Definition-IMSindex}, the accuracy of the interference ball model $S_{\beta,\xi}\left(\mathrm{IBM}\|\mathrm{PhyM} \right)$ is derived. A simple extension of our analysis gives the accuracy index when $d_0$ is a random variable. Recall that the purpose of this section is to illustrate only the utility of our index, and investigating more practical system models is a subject of our future work; see for instance~\cite{jiang2016simplified}.

\begin{result}[Perfect Interference Ball Model]\label{prop: perfect-IBM}
For any constant $0 \leq \xi \leq 1$ and any $\beta$, \newline $S_{\beta,\xi}\left(\mathrm{IBM}\|\mathrm{PhyM} \right) \to 1$ as $r_{\mathrm{IBM}} \to \infty$.
\end{result}

\begin{IEEEproof}
We know that $p_{\mathrm{fa}}^{{\mathrm{IBM}} \mid {\mathrm{PhyM}}} = 0$ for any constant $0 \leq \xi \leq 1$ and any $\beta$. Moreover, as $r_{\mathrm{IBM}}$ increases, $\Pr \left[\gamma^{\mathrm{IBM}} < \beta \right]$ tends to $\Pr \left[\gamma^{\mathrm{PhyM}} < \beta \right]$. Considering~\eqref{eq: miss-detection-uWave-Rayleigh-1}, $P_{\mathrm{md}}^{{\mathrm{IBM}} \mid {\mathrm{PhyM}}}$ asymptotically goes to zero as $r_{\mathrm{IBM}} \to \infty$. With zero false alarm and asymptotically zero miss-detection probabilities, the proof is concluded from~\eqref{eq: Definition-IMSindex}.
\end{IEEEproof}

Result~\ref{prop: perfect-IBM} indicates that the IBM becomes more accurate with higher $r_{\mathrm{IBM}}$, and it can be arbitrary accurate for sufficiently large $r_{\mathrm{IBM}}$. The price, however, is more complicated IBM as its approximations at a receiver demands coordination with more interferers.\footnote{Note that for special settings of this section, considering the impact of all interferers (PhyM) simplifies the analysis. However, this does not hold in general, e.g., if we change the spatial distribution of the interferers to a determinantal point process.} Also, negotiation with other transmitters (e.g., for MAC layer design) within this larger $r_{\mathrm{IBM}}$ becomes more challenging in terms of power consumption, signaling overhead, delay, and processing overhead.

\subsection{Accuracy of the Protocol Model}\label{subsec: example-1-PRM}
We now consider the PRM and first note that
\begin{align}\label{eq: false-alarm-prop2}
p_{\mathrm{fa}}^{{\mathrm{PRM}} \mid {\mathrm{PhyM}}} = 1 - \frac{\left( 1 - \Pr \left[\gamma^{\mathrm{PRM}} < \beta \right] \right) \hspace{-1mm} \left( 1- \Pr \left[\gamma^{\mathrm{PhyM}} < \beta \mid \gamma^{\mathrm{PRM}} \geq \beta \right]\right)}{1 - \Pr \left[\gamma^{\mathrm{PhyM}} < \beta \right]} \:,
\end{align}
and that
{\begin{align}\label{eq: misdetection-prop-2}
p_{\mathrm{md}}^{{\mathrm{PRM}} \mid {\mathrm{PhyM}}} = \frac{ \left(1 -  \Pr \left[\gamma^{\mathrm{PRM}} < \beta \right] \right) \Pr \left[\gamma^{\mathrm{PhyM}} < \beta \mid \gamma^{\mathrm{PRM}} \geq \beta \right]}{\Pr \left[\gamma^{\mathrm{PhyM}} < \beta \right]} \:.
\end{align}}
In the last two equations, note that $\Pr[\gamma^{\mathrm{PhyM}} < \beta ]$ is derived in~\eqref{eq: PhyM1-1}.
In the following, we derive $\Pr[\gamma^{\mathrm{PRM}} < \beta ]$ and $\Pr \left[\gamma^{\mathrm{PhyM}} < \beta \mid \gamma^{\mathrm{PRM}} \geq \beta \right]$.

Event $\gamma^{\mathrm{PRM}} < \beta$ occurs if there is at least one interferer inside $\mathcal{B}(2\pi,0,r_{\mathrm{PRM}})$. As $\mathcal{I}$ is a homogenous Poisson point process with intensity $ \lambda_t$, we have
\begin{equation}\label{eq: PRM-1}
\Pr \left[\gamma^{\mathrm{PRM}} < \beta \right] = 1 - \exp\left\{-\lambda_t \pi r_{\mathrm{PRM}}^2\right\} \:.
\end{equation}
In~Appendix~A, we have also derived
\begin{align}\label{eq: PRM1-2}
\hspace{-2mm}\Pr[\gamma^{\mathrm{PhyM}} < \beta \mid \gamma^{\mathrm{PRM}} \geq \beta]  &= 1 - \exp\Vast\{ \frac{- \sigma \beta d_{0}^{\alpha}}{pc} - \pi \lambda_t \, \mathbf{E}_h \vast[ - r_{\mathrm{PRM}}^2 \left( 1 - e^{- \beta d_{0}^{\alpha} h r_{\mathrm{PRM}}^{-\alpha}} \right) \vast. \Vast. \nonumber \\[-5mm]
& \hspace{-8mm} \Vast. \vast.
+ \left( \beta d_{0}^{\alpha} h \right)^{2/\alpha} \Gamma\left( 1 - \frac{2}{\alpha} \right)- \left( \beta d_{0}^{\alpha} h \right)^{2/\alpha} \Gamma\left( 1 - \frac{2}{\alpha}, \beta d_{0}^{\alpha} h r_{\mathrm{PRM}}^{-\alpha} \right) \vast] \Vast\}  \:.
\end{align}
Substituting~\eqref{eq: false-alarm-prop2}--\eqref{eq: PRM1-2} into~\eqref{eq: Definition-IMSindex}, we can find $S_{\beta,\Pr \left[ \gamma^{\mathrm{PhyM}} \geq \beta \right]}\left(\mathrm{PRM}\|\mathrm{PhyM} \right)$ for Rayleigh fading channel with omnidirectional transmission/reception.

\begin{result}[Miss-detection--False Alarm Tradeoff]\label{result: Pmd-Pfa-tradeoff}
Consider the protocol model of interference with Rayleigh fading channel. Increasing the interference range $r_{\mathrm{PRM}}$ reduces the false alarm probability and increases the miss-detection probability. Decreasing the interference range increases the false alarm probability and reduces the miss-detection probability.
\end{result}

\begin{IEEEproof}
$\Pr \left[\gamma^{\mathrm{PRM}} < \beta \right]$ is a strictly increasing function of $r_{\mathrm{PRM}}$, see~\eqref{eq: PRM-1}.
Considering the equations of the false alarm and miss-detection probabilities given in~\eqref{eq: false-alarm-prop2} and~\eqref{eq: misdetection-prop-2}, the proof concludes.
\end{IEEEproof}

\begin{result}[Asymptotic Accuracy of the Protocol Model]\label{result: zer-error}
Consider Equations~\eqref{eq: Definition-IMSindex} and~\eqref{eq: false-alarm-prop2}--\eqref{eq: PRM-1}. For any $0 \leq \xi \leq 1$ and any $\beta > 0$, we have the following asymptotic results:
\begin{alignat*}{3}
r_{\mathrm{PRM}} \to a , a \to 0 \hspace{1mm} &\Rightarrow &&\hspace{2mm} p_{\mathrm{fa}}^{{\mathrm{PRM}} \mid {\mathrm{PhyM}}} \to 0 \:, \hspace{2mm} p_{\mathrm{md}}^{{\mathrm{PRM}} \mid {\mathrm{PhyM}}} \to 1 \:, \hspace{2mm}  S_{\beta,\xi}\left(\mathrm{PRM}\|\mathrm{PhyM} \right) \to \xi
\:. \\ \nonumber \\[-2mm]
r_{\mathrm{PRM}} \to \infty \hspace{1mm} &\Rightarrow && \hspace{2mm} p_{\mathrm{fa}}^{{\mathrm{PRM}} \mid {\mathrm{PhyM}}} \to 1 \:, \hspace{2mm} p_{\mathrm{md}}^{{\mathrm{PRM}} \mid {\mathrm{PhyM}}} \to 0\:,  \hspace{2mm}   S_{\beta,\xi}\left(\mathrm{PRM}\|\mathrm{PhyM} \right) \to 1 - \xi
\:.
\end{alignat*}
\end{result}
Result~\ref{result: zer-error} further confirms the tradeoff between the miss-detection and false alarm probabilities.

\subsection{Numerical Illustrations}\label{sec: NumericalReults1}
To illustrate the accuracy index in Scenario~1 with Monte Carlo simulation, we consider a spatial Poisson network of interferers and obstacles with density $\lambda_t$ and $\lambda_o$ per unit area. Length of the typical link is $d_0 = 20$~m. We simulate a traditional outdoor microwave network~\cite{di2014stochastic} with average attenuation $c=22.7$~dB at the reference distance $a=1$~m, path-loss index $\alpha=3.6$, and noise power $\sigma = -111$~dBm (around 2~MHz bandwidth). We consider $p=20$~dBm transmission power and $\beta = 5$~dB minimum SINR threshold. For the ease of illustration, we define the notion of the \emph{average inter-transmitter distance} as $d_t = 1/\sqrt{\lambda_t}$. This distance directly relates to the inter-site distance in cellular networks, and also shows the transmitter density in a network.

Fig.~\ref{fig: uWave_Omni_Rayleigh__r} illustrates the impact of the interference range on the accuracy of both IBM and PRM under Scenario~1. From Fig.~\ref{subfig: uWave_Omni_Rayleigh__r_PfaPmd}, increasing $r_{\text{PRM}}$ increases $p_{\mathrm{fa}}^{{\mathrm{PRM}} \mid {\mathrm{PhyM}}}$ and reduces $p_{\mathrm{md}}^{{\mathrm{PRM}} \mid {\mathrm{PhyM}}}$, highlighted as the tradeoff between the miss-detection and false alarm probabilities in Result~\ref{result: Pmd-Pfa-tradeoff}. This tradeoff may lead to increment (see $d_t = 30$) or decrement (see $d_t = 80$) of the accuracy index of the PRM with the interference range. The IBM has zero false alarm probability, not depicted in Fig.~\ref{subfig: uWave_Omni_Rayleigh__r_PfaPmd} for sake of clarity of the figure. Moreover, as stated in Result~\ref{prop: perfect-IBM}, $p_{\mathrm{md}}^{{\mathrm{IBM}} \mid {\mathrm{PhyM}}}$ decreases with $r_{\text{PRM}}$, leading to a more accurate IBM, as can be confirmed in Fig.~\ref{subfig: uWave_Omni_Rayleigh__r_IMA}. Note that with the same transmitter density and interference range, the PRM has lower miss-detection probability than the IBM; however, better false alarm performance of the IBM leads to less errors in detecting outage events and therefore higher accuracy index, see Fig.~\ref{subfig: uWave_Omni_Rayleigh__r_IMA}. The TIM, not depicted in the figure, has a very high accuracy in all simulations. In particular, with $\varepsilon = -130$~dB, its accuracy is about 0.99. However, the corresponding TIM considers many interferers inside an irregular geometrical shape, which substantially decreases the tractability of  the resulting interference model.

\begin{figure}[t!]
  \centering
  \begin{subfigure}{0.49\columnwidth}
    \centering
    \includegraphics[width=\columnwidth]{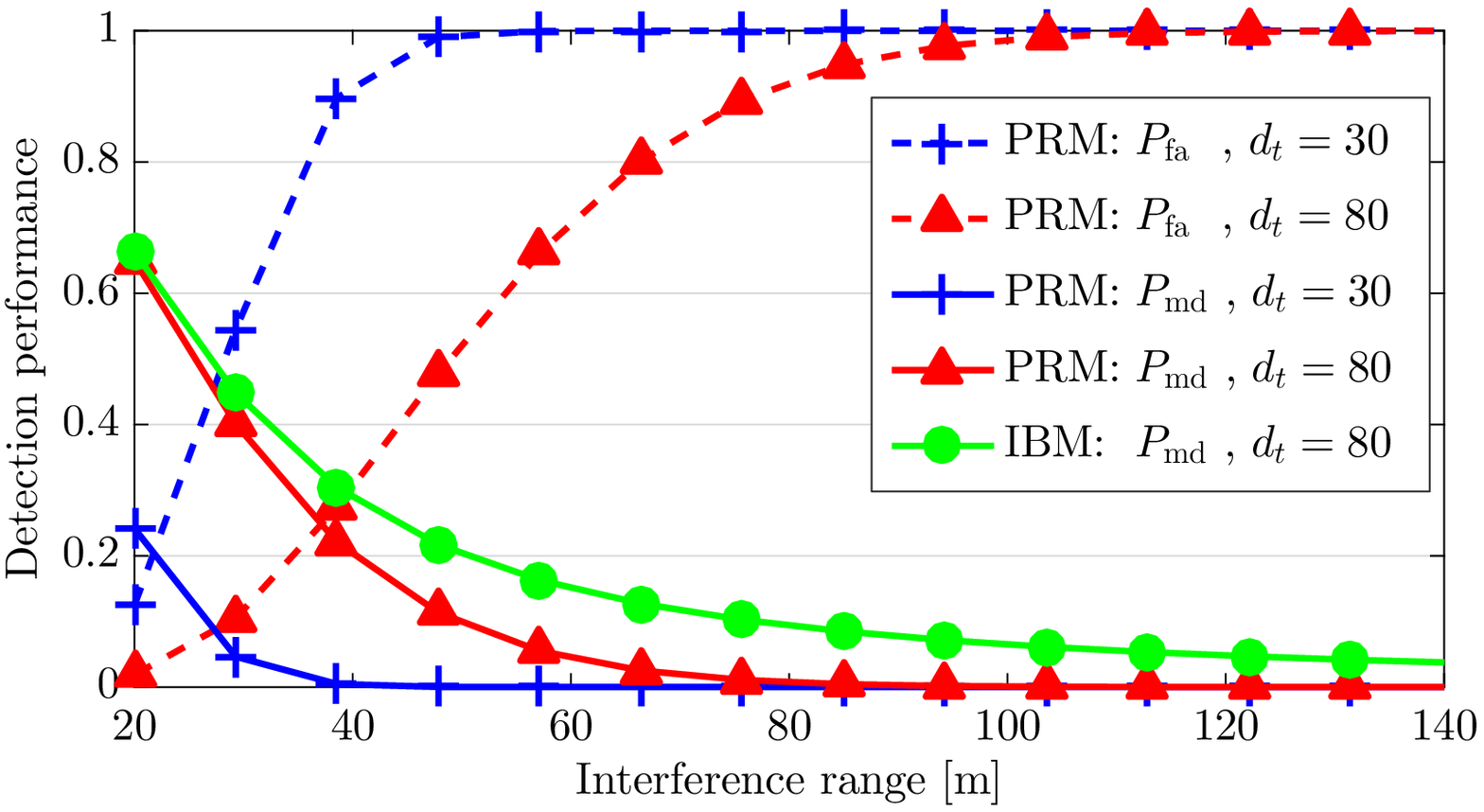}
    \caption{Error probabilities}
    \label{subfig: uWave_Omni_Rayleigh__r_PfaPmd}
  \end{subfigure}
  \begin{subfigure}{0.49\columnwidth}
       \centering
  \includegraphics[width=\columnwidth]{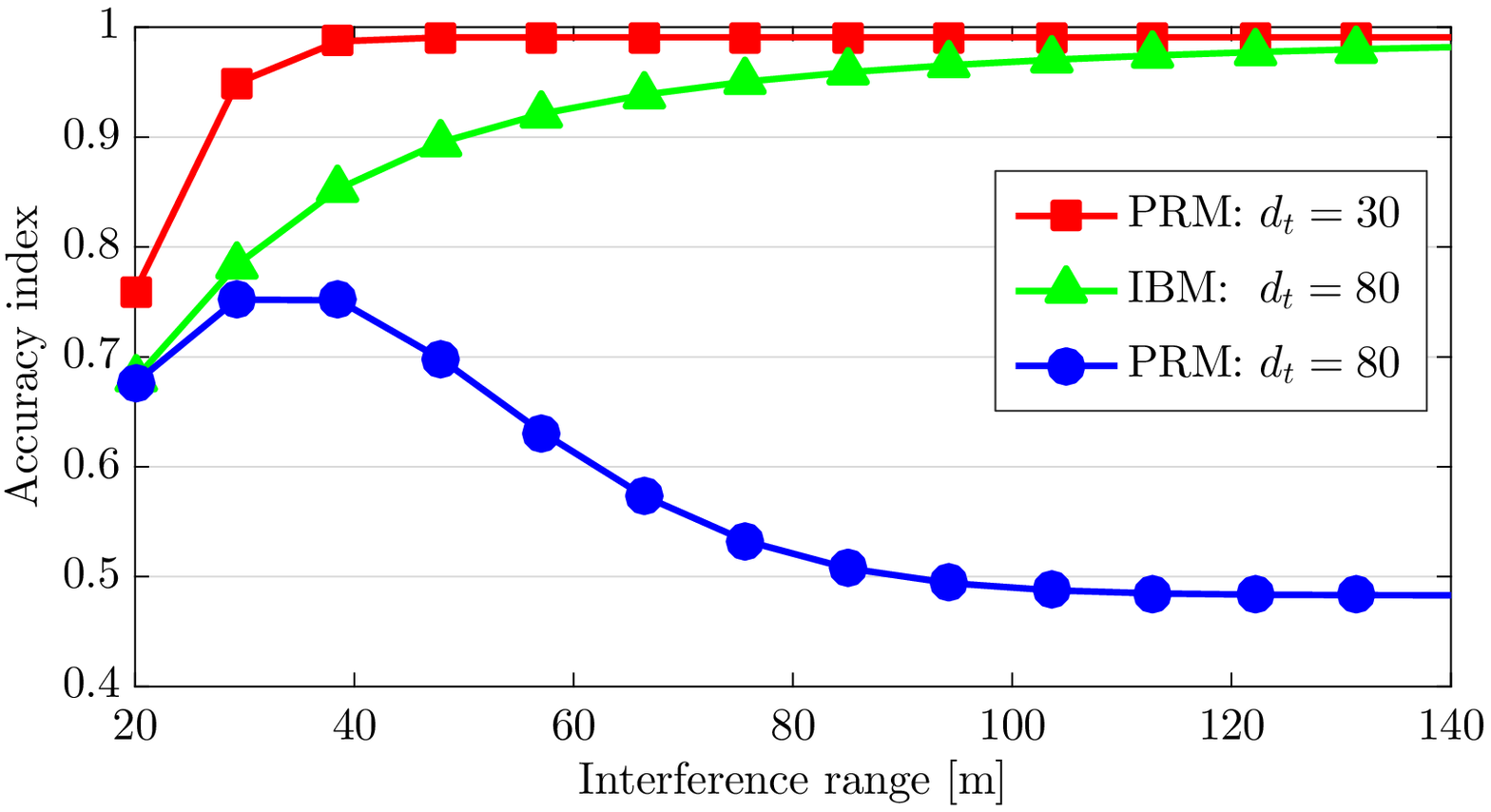}
    \caption{Accuracy index}
    \label{subfig: uWave_Omni_Rayleigh__r_IMA}
    \end{subfigure}
  \caption{Impact of the interference range on the accuracy of interference models under Rayleigh fading channel and omnidirectional communications.}
  \label{fig: uWave_Omni_Rayleigh__r}
\end{figure}
Fig.~\ref{fig: uWave_Omni_Rayleigh__lambdaT} shows the accuracy of the IBM and PRM under Scenario~1 against the average inter-transmitter distance. Again, we can observe an enhancement in the accuracy of the IBM with $r_{\text{IBM}}$, whereas the accuracy index of the PRM shows a complicated behavior as a function of $r_{\text{PRM}}$. By adopting the optimal $r_{\text{PRM}}$ that maximizes the accuracy index, as shown in Fig.~\ref{subfig: PRM_uWave_Omni_Rayleigh__lambdaT}, we can maintain a good performance for the PRM. Both interference models are very accurate at extremely dense transmitter deployments. The main reason is the very high interference level ($\xi = \Pr \left[ \gamma^{\text{PhyM}} \geq \beta \right]$ is almost 0 in this case), implying that the accuracy index is determined only by the miss-detection probability. Increasing the transmitter density through reducing $d_t$ decreases the miss-detection probability for both IBM and PRM, see Fig.~\ref{subfig: uWave_Omni_Rayleigh__r_PfaPmd}, improving their accuracy. For ultra sparse transmitter deployments, again, both interference models work accurately, as $\xi$ goes to 1 in this case and therefore only the false alarm probability determines the accuracy index. This probability is zero for the IBM, and it gets smaller values (asymptotically zero) for the PRM with higher $d_t$, see Fig.~\ref{subfig: uWave_Omni_Rayleigh__r_PfaPmd}. Finally, the TIM with $\varepsilon = -130$~dB, not shown in Fig.\ref{fig: uWave_Omni_Rayleigh__lambdaT}, has a very high accuracy in modeling the interference. Its accuracy for the same ranges of $d_t$ is higher than 0.98.

\begin{figure}[!t]
	\centering
    \begin{subfigure}[t]{0.49\columnwidth}
       \centering
		\includegraphics[width=\columnwidth]{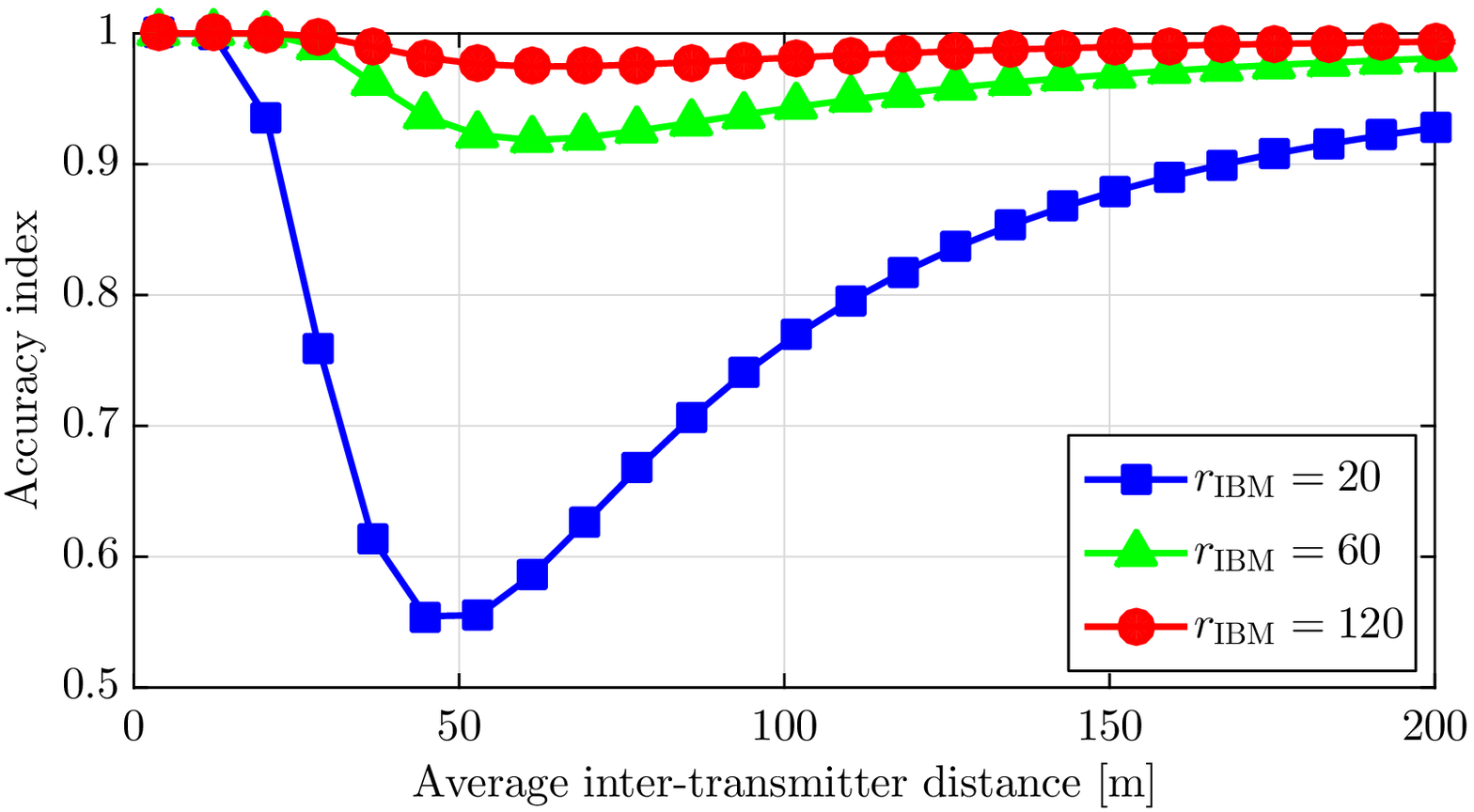}
	   \caption{Interference ball model}\label{subfig: IBM_uWave_Omni_Rayleigh__lambdaT}
    \end{subfigure}
    \begin{subfigure}[t]{0.49\columnwidth}
       \centering
       \includegraphics[width=\columnwidth]{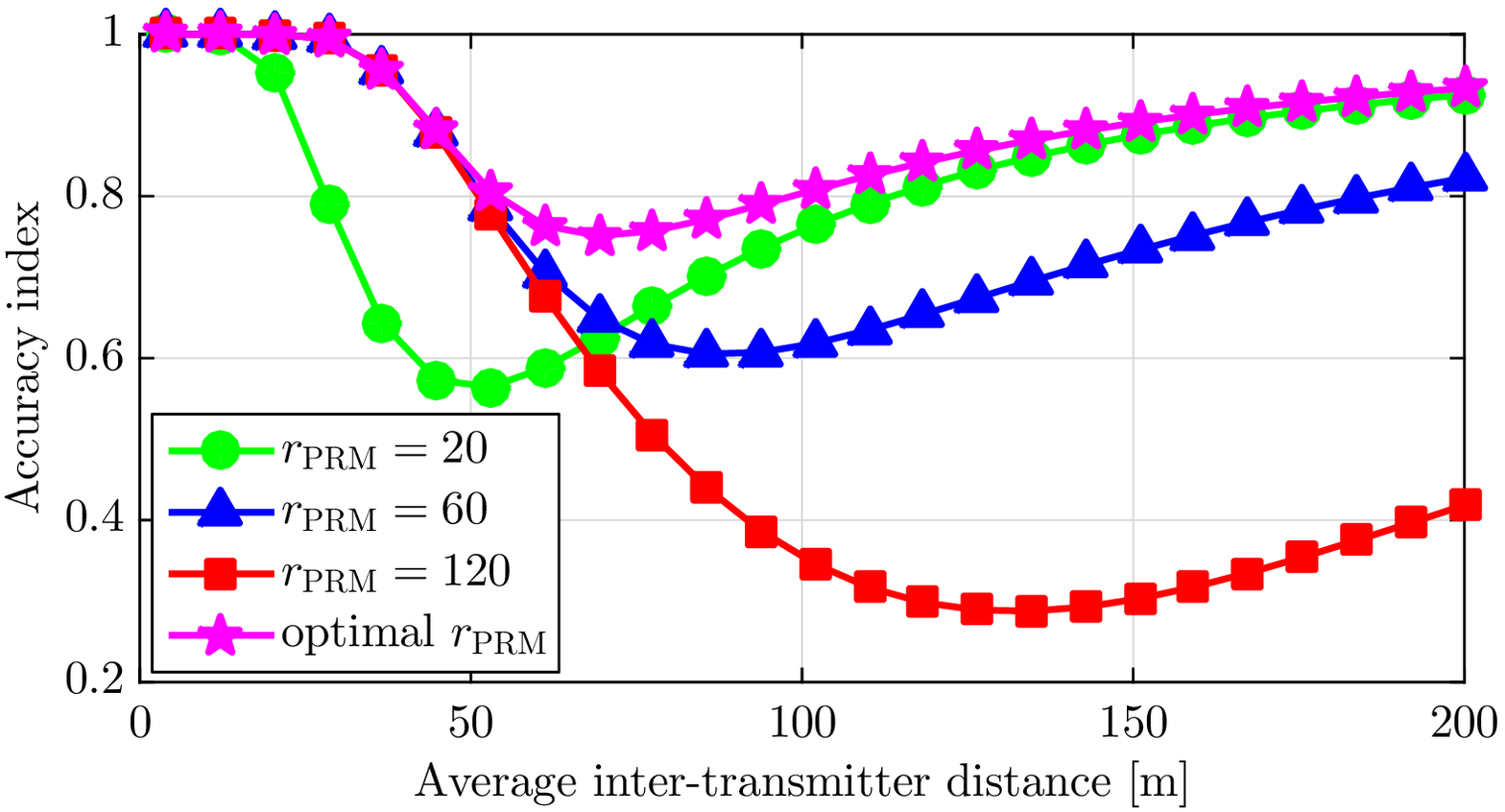}
	   \caption{Protocol model of interference}
       \label{subfig: PRM_uWave_Omni_Rayleigh__lambdaT}
    \end{subfigure}	
	
    \caption{Impact of transmitter density on the accuracy of the interference models under Rayleigh fading channel and omnidirectional communications. The accuracy of the TIM with $\varepsilon = -130$~dB is higher than 0.98.}
	\label{fig: uWave_Omni_Rayleigh__lambdaT}
\end{figure}

\begin{figure}[!t]
	\centering
	\includegraphics[width=0.75\columnwidth]{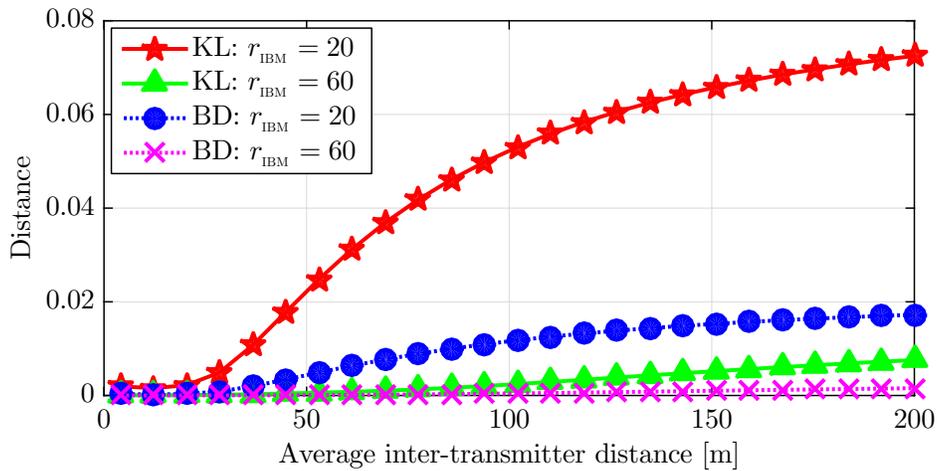}

    \caption{The KL divergence (labeled by ``KL'') of the distribution of $\gamma^{\mathrm{IBM}}$ from that of $\gamma^{\mathrm{PhyM}}$ and their Bhattacharyya distance (labeled by ``BD'') corresponding to the accuracy index values of Fig.~\ref{subfig: IBM_uWave_Omni_Rayleigh__lambdaT}. Lower values translate into higher similarity between two distributions.}
	\label{fig: KL_BC_Scenario1}
\end{figure}
Fig.~\ref{fig: KL_BC_Scenario1} shows the KL divergence of $f_{\gamma^{\mathrm{IBM}}}(x)$ from $f_{\gamma^{\mathrm{PhyM}}}(x)$ and also their Bhattacharyya distance for the same setting of Fig.~\ref{subfig: IBM_uWave_Omni_Rayleigh__lambdaT}, where lower values translates into higher accuracy of the IBM. From this figure, both the KL divergence and the Bhattacharyya distance can identify higher accuracy of the IBM with $r_{\text{IBM}}=60$~m. However, they both fail to show that the performance of IBM with $r_{\text{IBM}}=20$~m converges to that with $r_{\text{IBM}}=60$~m once the network gets sparser. Moreover, calculating these measures entails almost the same mathematical/numerical complexity as our similarity index. Due to these reasons, we investigate only our accuracy index for the rest of the paper, though one may incorporate those metrics in our proposed interference model similarity analysis framework.

Fig.~\ref{fig: IBM_uWave_Omni_Rayleigh__beta} illustrates the accuracy index against the SINR threshold. Increasing the SINR threshold generally increases the sensitivity of the interference model to any approximation error in $\mathrm{x}$.
\begin{figure}[!t]
  \centering
  \includegraphics[width=0.75\columnwidth]{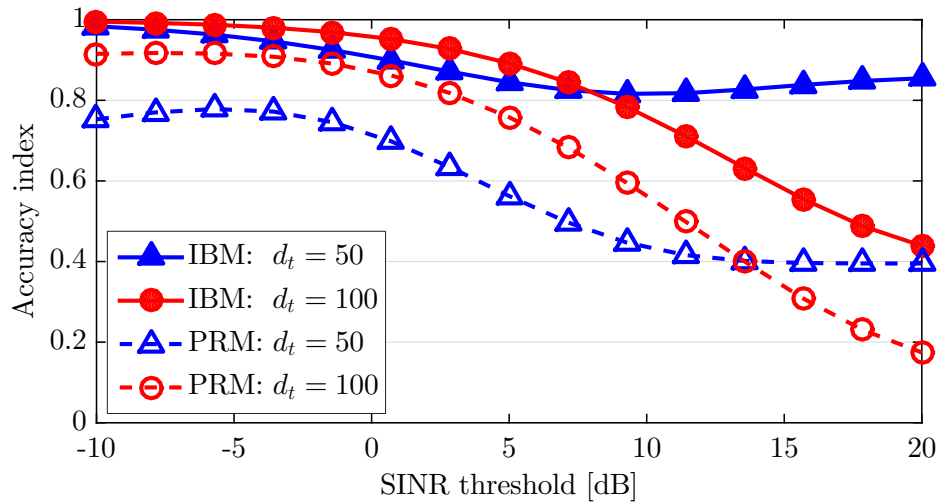}

  \caption{Impact of the SINR threshold on the accuracy of interference models under Rayleigh fading channel and omnidirectional communications.}
  \label{fig: IBM_uWave_Omni_Rayleigh__beta}
\vspace{-7mm}
\end{figure}

\section{Example Scenario~2: Rayleigh Fading Channel, Directionality, and Obstacles}\label{sec: Rayleigh-Blockage}
In this section, we analyze the accuracy of IBM and PRM in modeling a wireless network with Rayleigh fading channels, where all transmitters and receivers use directional communications to boost the link budget and to reduce multiuser interference. We also consider impenetrable obstacles. The application areas of this scenario include modeling and performance evaluation of mmWave networks, where directional communication is inevitable and extreme penetration loss due to most of the solid materials (e.g., 20--35~dB due to the human body~\cite{Rangan2014Millimeter}) justifies the impenetrable obstacle assumption. In Section~\ref{sec: OtherComponents}, we will comment on the impact of assuming impenetrable obstacles on the accuracy of the interference model.

Note that the interference is not the primary limitation of mmWave networks specially if we take an average over all possible realizations of a random topology~\cite{di2014stochastic,Shokri2015Transitional}. However, even if mmWave networks are noise-limited in a statistical sense (that is, taking an average of the interference over some time or some topologies), there are significant realizations of network topologies at given times where some transmitters can cause strong interference. We cannot use noise-limited arguments, which are valid over some time horizons, when we have to optimize in real-time resource allocations or routing. In the following two sections, we show that special characteristics of mmWave networks, such as blockage and deafness, can be exploited to substantially simplify the interference model, so as to develop efficient scheduling and routing algorithms, which may otherwise be impossible. In fact, our results provide, for the first time, mathematical justifications for the use of simpler interference models in mmWave networks, as extensively done in the literature~\cite{Singh2011Interference,Singh2010Distributed,an2008directional,son2012frame,petrov2017interference,garcia2015analysis,Yu2016Distributed,Stahlbuhk2016Topology}.

We assume a homogenous Poisson network of interferers as in Section~\ref{sec: Rayleigh-noBlockage}. If there is no obstacle on the link between transmitter $i$ and the typical receiver located at the origin, we say that transmitter $i$ has line-of-sight (LoS) condition with respect to the typical receiver, otherwise it is in non-LoS condition. We assume that transmitter of every link is spatially aligned with its intended receiver, so there is no beam-searching phase~\cite{Shokri2015Beam}.
We model the antenna pattern by an ideal sector model~\cite{di2014stochastic}, where the antenna gain is a constant in the main lobe and another smaller constant in the side lobe. We assume the same operating beamwidth $\theta$ for all devices in both transmission and reception modes. Then, the antenna gain for each transmitter/receiver is~\cite[Equation~(3)]{Shokri2015Beam}
\begin{equation}\label{eq: SectorAntennaGains}
  \begin{cases}
    \dfrac{2\pi - \left(2\pi-\theta\right)z}{\theta}, & \mbox{inside the main lobe} \\
    z, & \mbox{inside the side lobe}\:,
  \end{cases}
\end{equation}
where $0\leq z \ll 1$ is the side lobe gain. For mathematical tractability, we assume negligible side lobe gain (i.e., $z=0$) throughout this section, and numerically assess the impact of this simplification in Section~\ref{sec: OtherComponents}.

Consider the link between transmitter $i$ and receiver $j$ with distance $d_{ij}$. It is shown that with a random number of obstacles, each having random location and size, this link is in the LoS condition with probability $e^{-\epsilon \lambda_o d_{ij}}$, where $\lambda_o$ is the intensity of the obstacles and $\epsilon$ is a constant value that depends on the average size of obstacles in the environment~\cite{TBai2014Blockage}. Due to the exponential decrease of the LoS probability with the link length (also see~\cite[Fig.~4]{Rappaport2015wideband}), very far interferers are most likely blocked. For mathematical simplicity, we assume independent LoS conditions among the typical receiver and all other transmitters, and also impenetrable obstacles. Nonetheless, the following analysis can be extended for more realistic blockage models, introduced in~\cite{Shokri2015Transitional}. Notice that we are using this simplified model to investigate the effects of directionality and blockage on the accuracy of the interference models and to characterize fundamental properties of the proposed accuracy index. The exact value of the accuracy index with a more realistic mmWave channel can be readily numerically calculated under any system model, as we highlight in the next sections.

%

To evaluate the accuracy of IBM and PRM, we first notice that an intended transmitter can cause a significant interference contribution to the typical receiver if: (a) the typical receiver is inside its main lobe, (b) it has LoS condition with respect to the typical receiver, and (c) it is inside the main lobe of the typical receiver.
Due to random deployment of the transmitters and receivers, the probability that the typical receiver locates inside the main lobe of a transmitter is $\theta/2 \pi$. Moreover, we have independent LoS events among the typical receiver and individual transmitters. Therefore, the interferers for which conditions~(a)--(b) hold follow an inhomogeneous Poisson point process $\mathcal{I}$ with intensity of $\lambda_I \left( r \right) = \lambda_t \theta e^{-\epsilon\lambda_o r} / 2\pi$ at radial distance $r$. Condition~(c) reduces the angular region that a potential interferer should be located to contribute in the interference observed by the typical receiver.
We note that $\mathcal{I} \cap \mathcal{B}(\theta,0,r_{\mathrm{PRM}})$ is the set of potential interferers inside the vulnerable region of the PRM, shown by red triangles in Fig.~\ref{fig: IntRegion}, and $\mathcal{I} \cap \mathcal{B}(\theta,r_{\mathrm{PRM}},\infty)$ shows the set of potential interferers outside that region, shown by green circles in Fig.~\ref{fig: IntRegion}. Also, $\mathcal{I} \cap \mathcal{B}(\theta,0,r_{\mathrm{IBM}})$ is the set of potential interferers for IBM (near-field interferers).
\begin{figure}[!t]
  \centering
  \includegraphics[width=0.32\columnwidth]{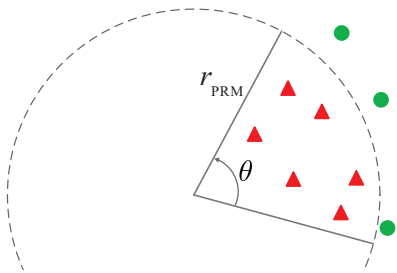}\\

\vspace{1mm}
  \caption{Illustration of the vulnerable area.}\label{fig: IntRegion}
\vspace{-7mm}
\end{figure}
%

\subsection{Impact of Directionality and Blockage}
Before deriving the accuracy of IBM and PRM, we first evaluate the impact of directionality and blockage on the number of the interferers. We define by $\Lambda_{\mathcal{B}(\theta,0,R)}$ the measure of the region $\mathcal{B}(\theta,0,R)$, i.e., the average number of interferers inside the region. We have
\begin{equation}\label{eq: measure-of-region}
\Lambda_{\mathcal{B}(\theta,0,R)} \hspace{-0.5mm} =  \hspace{-0.5mm}  \theta  \hspace{-1mm} \int_{0}^{R}{ \hspace{-2mm}  \lambda_I (r) r\, \mathrm{d}r}  \hspace{-0.5mm} =  \hspace{-0.5mm} \frac{\theta^2 \lambda_t}{2\pi \epsilon^{2} \lambda_{o}^{2}} \Big( 1 - \left( 1 + \epsilon \lambda_o R \right)e^{- \epsilon \lambda_o R}\Big) \:.
\end{equation}
Then, for any real $R>0$, the number of potential interferer inside the region $\mathcal{B}(\theta,0,R)$, denoted by $N_{\mathcal{B}(\theta,0,R)}$, is a Poisson random variable with probability mass function
\begin{equation}\label{eq: number-of-points}
  \Pr[N_{\mathcal{B}(\theta,0,R)}=n] = e^{-\Lambda_{\mathcal{B}(\theta,0,R)}} \frac{\left(\Lambda_{\mathcal{B}(\theta,0,R)} \right)^{n}}{n!} \:.
\end{equation}

\begin{result}[Impact of Directionality]\label{result: impact-of-directionality}
Consider~\eqref{eq: measure-of-region}, and let ${\epsilon\lambda_o \to 0}$. The average number of potential interferers converges to
\begin{equation}
\frac{\theta^2 \lambda_t}{4\pi} R^2 = \left(\frac{\theta}{2\pi} \lambda_t\right) \left( \frac{\theta}{2}R^2\right) \:.
\end{equation}
\end{result}
To interpret Result~\ref{result: impact-of-directionality}, with no obstacle in the environment (${\epsilon\lambda_o \to 0}$), we will have a homogenous Poisson network of interferers with density $\lambda_t \theta/ 2 \pi$. Therefore, the average number of interferers over $\mathcal{B}(\theta,0,R)$ is the product of the density per unit area and the area of $\mathcal{B}$, which is $\theta R^2/2$. It can be concluded that adopting narrower beams reduces the average number of potential interferers within a certain distance $R$; however, it still tends to infinity almost surely as $R \to \infty$.

\begin{result}[Impact of Blockage]\label{result: impact-of-obstacle}
Consider~\eqref{eq: measure-of-region}, and let ${R \to \infty}$. The average number of potential interferers converges to
\begin{equation}\label{eq: avg-number-interferers}
\frac{\theta^2 \lambda_t}{2\pi \epsilon^{2} \lambda_{o}^{2}} \:,
\end{equation}
which is less than infinity almost surely if $\epsilon\lambda_o > 0$.
\end{result}

Result~\ref{result: impact-of-obstacle} implies that any receiver observes a finite number of potential interferers almost surely if there is a non-negligible blockage. This unique feature holds for the mmWave bands, as most of the obstacles can severely attenuate the signals.\footnote{In the conventional microwave systems where the transmission is less sensitive to blockage, the number of potential interferers is almost surely infinite, as highlighted in Result~\ref{result: impact-of-directionality}.} Therefore, not only the farther transmitters will contribute less on the aggregated interference (due to higher path-loss), but they will be also thinned by directionality and blockage such that only a \emph{finite} number of spatially close transmitters can cause non-negligible interference to any receiver. Note that, these fewer interferers may still cause strong interference, if they are located very close to the receiver. The point is that the thinning process due to directionality and blockage makes the SINR distribution under PhyM closer to that of the IBM, which considers only the near-field interferers. To elaborate more, we characterize the average number of far-field interferers in the following.

\begin{prop}[Measure of Far-Field Interferers]\label{prop: measure-far-field}
Let $\theta$ be the operating beamwidth, $\lambda_t$ be the density of the transmitters, $\lambda_o$ be the density of the obstacles, and $\epsilon>0$ be a constant. Then, the average number of interferers located inside $\mathcal{B}(\theta,R,\infty)$ is
\begin{equation}\label{eq: measure-far-field}
\Lambda_{\mathcal{B}(\theta,R,\infty)} = \frac{\theta^2 \lambda_t}{2\pi \epsilon^{2} \lambda_{o}^{2}} \left( 1 + \epsilon \lambda_o R \right)e^{- \epsilon \lambda_o R} \:,
\end{equation}
and the probability of having no far-field interferer is
\begin{equation}\label{eq: AvgNoPoints-far-field}
\Pr[N_{\mathcal{B}(\theta,R,\infty)}=0] = e^{-\Lambda_{\mathcal{B}(\theta,R,\infty)}} \:.
\end{equation}
\end{prop}

\begin{IEEEproof}
To prove, we only need to compute $\int_{R}^{\infty}{\! \theta \lambda_I \left( r \right) r\, \mathrm{d}r}$, and~\eqref{eq: measure-far-field} follows. Moreover, by substituting $\Lambda_{\mathcal{B}(\theta,R,\infty)}$ into~\eqref{eq: number-of-points} with $n=0$, we conclude~\eqref{eq: AvgNoPoints-far-field}.
\end{IEEEproof}

From Proposition~\ref{prop: measure-far-field}, the average number of far-field interferers will be decreased exponentially with distance. Consequently, from~\eqref{eq: AvgNoPoints-far-field}, the probability of having no far-field interferers increases exponentially with the distance. Fig.~\ref{fig: FarFieldInterference} shows the probability of having at least one far-field interferer as a function of the distance. By defining any arbitrary minimum threshold $\kappa$ on this probability, we can find a distance $R_{\kappa}$ after which the probability of having far-field interferer(s) is arbitrarily close to 0 (less than $\kappa$). This suggests that by setting $r_{\mathrm{IBM}} = R_{\kappa}$, IBM can capture, at least, $(1-\kappa)$ fraction of the total number of interferers for any arbitrary small $\kappa$. Recall that the neglected interferers, if any, are far-field and their contributions to the total interference term are suppressed by the significant distance-dependent path-loss. All these facts result in the following conclusion:
\begin{result}\label{result: 6}
Directionality and blockage can substantially increase the accuracy of the interference ball model.
\end{result}
We can argue similar accuracy improvement in the PRM, as we numerically illustrate in the next subsections.

\begin{figure}[!t]
  \centering
  \includegraphics[width=0.75\columnwidth]{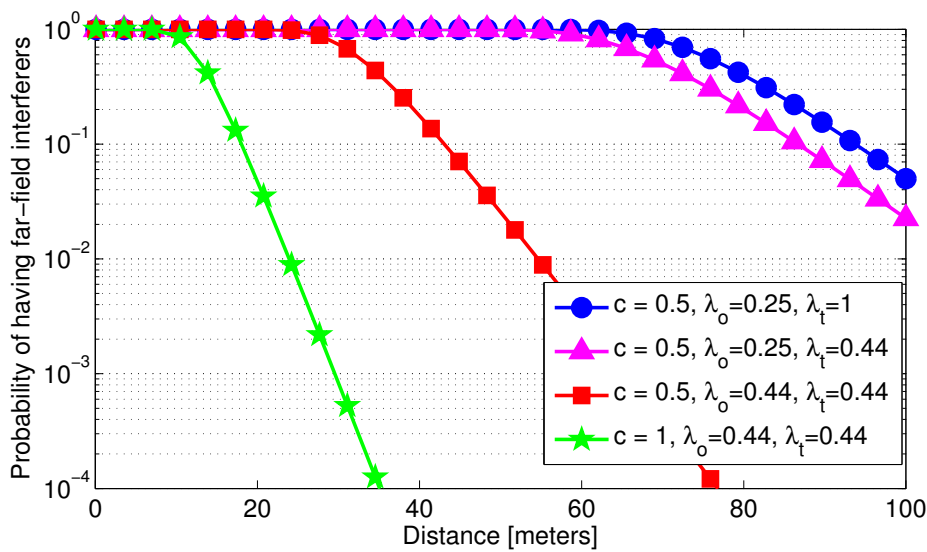}\\

  \caption{Probability of having at least one far-field interferer as a function of distance. Simulation parameters are similar to those of Fig.~\ref{fig: uWave_Omni_Rayleigh__r}.}
  \label{fig: FarFieldInterference}
\vspace{-7mm}
\end{figure}

\subsection{Accuracy of the Interference Ball Model}\label{subsec: example-2-IBM}
Assume $r_{\mathrm{IBM}} \geq a$ and $d_0 \geq a$.
Using similar claims as in Section~\ref{subsec: example-1-IBM}, it is straightforward to show $p_{\mathrm{fa}}^{{\mathrm{IBM}} \mid {\mathrm{PhyM}}} = 0$.
To find the miss-detection probability, in Appendix~B, we derive
\begin{align}\label{eq: IBM-3}
\Pr\left[\gamma^{\mathrm{IBM}} < \beta \right]  &= 1 - \exp\Vast\{ - \frac{\sigma \theta^2 \beta d_{0}^{\alpha}}{4 pc\pi^2} - \frac{\theta^2}{2 \pi} \lambda_t  \, \mathbf{E}_h \vast[\left( 1 - e^{- \beta d_{0}^{\alpha} h} \right) \left( \frac{1 - \left( \epsilon \lambda_o a+ 1 \right) e^{-\epsilon \lambda_o a}}{\epsilon^2 \lambda_{o}^{2}}\right) \vast. \Vast. \nonumber \\[-5mm]
& \hspace{58mm} \vast. \Vast. + \int_{a}^{r_{\mathrm{IBM}}} \! \left( 1 - e^{- \beta d_{0}^{\alpha} h r^{-\alpha}} \right) e^{-\epsilon\lambda_o r} r \, \mathrm{d}r \vast] \Vast\} \:,
\end{align}
and
\begin{align}\label{eq: PhyM-New1}
\Pr \left[\gamma^{\mathrm{PhyM}} < \beta \right] &= 1 - \exp\Vast\{ - \frac{\sigma \theta^2 \beta d_{0}^{\alpha}}{4 pc\pi^2} - \frac{\theta^2 \lambda_t}{2 \pi \epsilon^2 \lambda_{o}^{2}} \, \mathbf{E}_h \vast[ 1 - e^{- \beta d_{0}^{\alpha} h} \left( 1 - \left( \epsilon \lambda_o a+ 1 \right) e^{-\epsilon \lambda_o a}\right) \vast. \Vast. \nonumber \\[-5mm]
& \hspace{64mm} \Vast. \vast. - \epsilon^2 \lambda_{o}^{2} \int_{a}^{\infty} \! e^{- \beta d_{0}^{\alpha} h r^{-\alpha} -\epsilon\lambda_o r} r \, \mathrm{d}r \vast] \Vast\}  \:,
\end{align}
and substitute them into~\eqref{eq: miss-detection-uWave-Rayleigh-1}. Then, $S_{\beta,\xi}\left(\mathrm{IBM}\|\mathrm{PhyM} \right)$ can be found using~\eqref{eq: IBM-3}, \eqref{eq: PhyM-New1}, \eqref{eq: miss-detection-uWave-Rayleigh-1}, and then~\eqref{eq: Definition-IMSindex}. Similar to Remark~\ref{prop: perfect-IBM}, for any $0 \leq \xi \leq 1$, $S_{\beta,\xi}\left(\mathrm{IBM}\|\mathrm{PhyM} \right) \to 1$ as $r_{\mathrm{IBM}} \to \infty$.

\subsection{Accuracy of the Protocol Model}\label{subsec: example-2-PRM}
To derive the accuracy of the PRM, we need to derive $\Pr[\gamma^{\mathrm{PhyM}} < \beta ]$, $\Pr[\gamma^{\mathrm{PRM}} \leq \beta]$ and $\Pr[\gamma^{\mathrm{PhyM}} < \beta \mid \gamma^{\mathrm{PRM}} \geq \beta]$, and substitute them into~\eqref{eq: false-alarm-prop2} and~\eqref{eq: misdetection-prop-2}. $\Pr[\gamma^{\mathrm{PhyM}} < \beta ]$ is derived in~\eqref{eq: PhyM-New1}.

Event $\gamma^{\mathrm{PRM}} < \beta$ implies that ${\left| \mathcal{I} \cap \mathcal{B}(\theta,0,r_{\mathrm{PRM}}) \right| \geq 1}$, namely there is at least one potential interferer inside $\mathcal{B}(\theta,0,r_{\mathrm{PRM}})$. Considering~\eqref{eq: number-of-points}, the probability of this event is ${\Pr [N_{\mathcal{B}(\theta,0,r_{\mathrm{PRM}})} \geq 1]}$, thus
\begin{align}\label{eq: PRM-3}
\Pr \left[\gamma^{\mathrm{PRM}} < \beta \right] = 1 - \exp\Big\{-\Lambda_{\mathcal{B}(\theta,0,r_{\mathrm{PRM}})} \Big\} \:,
\end{align}
Event $\gamma^{\mathrm{PRM}} \geq \beta$ implies that there is no interferer inside $\mathcal{B}(\theta,0,r_{\mathrm{PRM}})$. Assuming $r_{\mathrm{PRM}} \geq a$, it is easy to find $\Pr[\gamma^{\mathrm{PhyM}} < \beta \mid \gamma^{\mathrm{PRM}} \geq \beta]$:
\begin{equation}\label{eq: PRM-4}
\Pr[\gamma^{\mathrm{PhyM}} < \beta \mid \gamma^{\mathrm{PRM}} \geq \beta] = 1 - \exp\Vast\{\hspace{-1mm}- \frac{\sigma \theta^2 \beta d_{0}^{\alpha}}{4 pc\pi^2} - \frac{\theta^2 \lambda_t}{2 \pi } \, \mathbf{E}_h \vast[\int_{r_{\mathrm{PRM}}}^{\infty}\hspace{-2mm} \left( 1 - e^{- \beta d_{0}^{\alpha} h r^{-\alpha}} \right) e^{-\epsilon\lambda_o r} r \, \mathrm{d}r \vast]\hspace{-2mm} \Vast\} .
\end{equation}
Substituting~\eqref{eq: PhyM-New1}--\eqref{eq: PRM-4} into~\eqref{eq: false-alarm-prop2} and~\eqref{eq: misdetection-prop-2} gives the accuracy index of the PRM. Note that Results~\ref{result: Pmd-Pfa-tradeoff} and~\ref{result: zer-error} hold here.

\subsection{Numerical Illustrations}\label{sec: NumericalResult2}
To numerically illustrate the accuracy index in Scenario~2, we use the same simulation environment of Section~\ref{sec: NumericalReults1}. We independently randomly mark some wireless link to be blocked by obstacles, with the exponential blockage probability with $\epsilon \lambda_o = 0.008$~\cite{TBai2014Blockage}. We then assume infinite penetration loss for the blocked links, and use the large scale LoS path loss model at 28~GHz~\cite[Table~I]{Akdeniz2014MillimeterWave}. System bandwidth is 1~GHz (noise power $\sigma = -84$~dBm). Without loss of generality, we assume $r_{\text{PRM}} = 40$~m and $r_{\text{IBM}} = 80$~m.

\begin{figure}[!t]
  \centering
  \includegraphics[width=0.7\columnwidth]{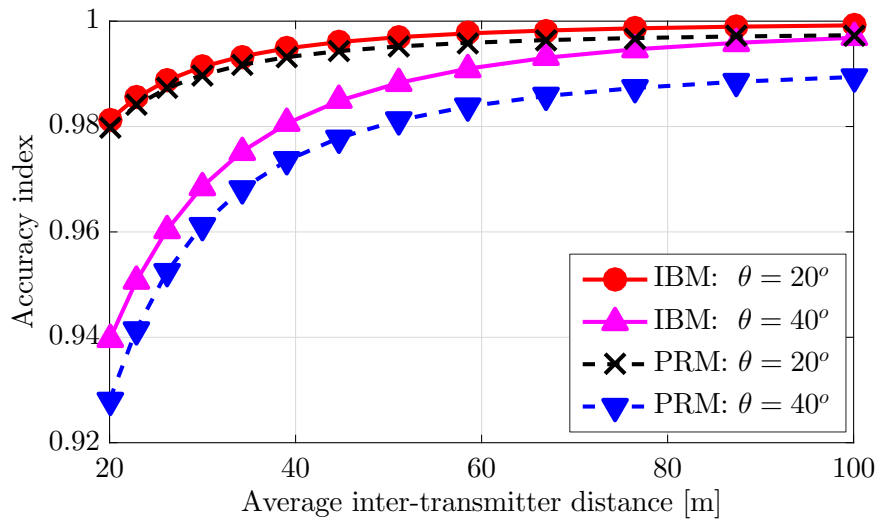}

  \caption{Accuracy of IBM and PRM under Rayleigh fading channel and directional communications with obstruction.}
  \label{mmWave_Direc_Deterministic__lambdaT2}
\vspace{-7mm}
\end{figure}
Fig.~\ref{mmWave_Direc_Deterministic__lambdaT2} illustrates the impact of the operating bandwidth and average inter-transmitter distance on the accuracy index of both IBM and PRM under Scenario~2. As expected, the IBM outperforms PRM. More importantly, directionality and blockage improve the accuracy of both interference models. We show in the following section that changing the underlying channel model from a Rayleigh fading model to a deterministic model further enhances their accuracies. Moreover, the accuracy of the TIM with ${\varepsilon = -130}$~dB, not depicted for the sake of clarity in Fig.~\ref{mmWave_Direc_Deterministic__lambdaT2}, is nearly 1 in our simulations. Notice that a simplified interference model (e.g., PRM, IBM, or TIM) may not be of sufficient accuracy for all range of parameters, still it is substantially improved by directionality and blockage, as highlighted by Results~\ref{result: impact-of-directionality}--\ref{result: 6}.

%

%

\section{Example Scenario~3: Deterministic Channel, Directionality, and Obstacles}\label{sec: Scenarios3}
In this section, we investigate how accurately the IBM and PRM can model a wireless network with directional communications, blockage, and deterministic wireless channel. The last assumption holds generally in mmWave networks, where sparse scattering characteristic of mmWave frequencies along with narrow beam operation makes the mmWave channel more deterministic compared to that of microwave systems with rich scattering environment and omnidirectional operation~\cite{Rappaport2015wideband}.

\subsection{Accuracy of the Interference Ball Model}\label{subsec: example-3-IBM}
Again, it is straightforward to show $p_{\mathrm{fa}}^{{\mathrm{IBM}} \mid {\mathrm{PhyM}}} = 0$. However, unlike previous cases, we cannot derive closed-form expression for the miss-detection probability, and consequently for the accuracy index. In Appendix~C, we have derived upper bounds on the miss-detection probability using the Chernoff bound.

\subsection{Accuracy of the Protocol Model}\label{subsec: example-3-PRM}
Again, deterministic wireless channel prohibits deriving closed-form expressions for the false alarm and miss-detection probabilities. Nevertheless, we can show that both Remarks~2 and~3 holds here. Moreover, we have the following result:
\begin{result}[Zero False Alarm Probability]\label{result: 7}
Under the deterministic channel model, the false alarm probability is zero for any $r_{\mathrm{PRM}} \leq \zeta^{-1/\alpha}$, where
\begin{equation}\label{eq: zeta}
\zeta =   \frac{d_{0}^{-\alpha}}{\beta} - \frac{\sigma}{pc}\left( \frac{\theta }{2 \pi} \right)^{2} \:.
\end{equation}
\end{result}
\begin{IEEEproof}
The SINR that the typical receiver experiences due to transmission of the intended transmitter and an unintended receiver located at distance $r_{\mathrm{PRM}}$ is
\begin{equation*}
\frac{p c \left( \frac{2 \pi}{\theta} \right)^2 d_{0}^{-\alpha} }{p c \left( \frac{2 \pi}{\theta} \right)^2 r_{\mathrm{PRM}}^{-\alpha} +\sigma} \:.
\end{equation*}
Comparing the SINR expression to $\beta$, we get that any interferer located at distance $r_{\mathrm{PRM}}$ less than
\begin{equation*}
\left( \frac{d_{0}^{-\alpha}}{\beta} - \frac{\sigma}{pc}\left( \frac{\theta}{2 \pi} \right)^{2} \right)^{-1/\alpha}  = \zeta^{-1/\alpha}
\end{equation*}
can cause packet loss at the typical receiver, namely ${\gamma^{\mathrm{PhyM}} < \beta}$.
Now, if we consider the general equation of the false alarm probability, we have
\begin{align*}
p_{\mathrm{fa}}^{{\mathrm{PRM}} \mid {\mathrm{PhyM}}}  = \Pr \left[\gamma^{\mathrm{PRM}} < \beta \mid \gamma^{\mathrm{PhyM}} \geq \beta \right]  = \frac{\Pr \left[\gamma^{\mathrm{PRM}} < \beta \right] \Pr \left[\gamma^{\mathrm{PhyM}} \geq \beta \mid \gamma^{\mathrm{PRM}} < \beta \right]}{1 - \Pr \left[\gamma^{\mathrm{PhyM}} < \beta \right]} \:.
\end{align*}
For $r_{\mathrm{PRM}} \leq \zeta^{-1/\alpha}$, $\Pr \left[\gamma^{\mathrm{PhyM}} \geq \beta \mid \gamma^{\mathrm{PRM}} < \beta \right] = 0$; since event $\gamma^{\mathrm{PRM}} < \beta$ implies that there is at least one interferer inside $\mathcal{B}(\theta,0,r_{\mathrm{PRM}})$. This interferer ensures $\gamma^{\mathrm{PhyM}} < \beta$, so $p_{\mathrm{fa}}^{{\mathrm{PRM}} \mid {\mathrm{PhyM}}} = 0$ for $r_{\mathrm{PRM}} \leq \zeta^{-1/\alpha}$.
\end{IEEEproof}

The following proposition characterizes bounds for the accuracy index for the Example Scenario~3 (mmWave networks):
\begin{prop}\label{prop: bounds-PRM}
For $\xi = \Pr \left[ \gamma^{\mathrm{PhyM}} \geq \beta \right]$ and any $0 < r_{\mathrm{PRM}} \leq \zeta^{-1/\alpha}$, we have
\begin{equation*}
\Pr \left[\gamma^{\mathrm{PRM}}  < \beta \right] \leq S_{\beta,\xi}\left(\mathrm{PRM}\|\mathrm{PhyM} \right) \leq 1 \:,
\end{equation*}
where $\Pr \left[\gamma^{\mathrm{PRM}}  < \beta \right]$ is given in~\eqref{eq: PRM-3}.
\end{prop}
We have provided a proof for this proposition along with other bounds in Appendix~C. We have the following scaling law results:
\begin{result}[Scaling laws for the PRM]
The following scaling laws are implied by Proposition~\ref{prop: bounds-PRM} and inequality $e^{x} \geq 1+x$ for any $x\geq 0$:
\begin{itemize}
  \item \emph{Scaling with $\theta$:} For any constant $r_{\mathrm{PRM}}$ no larger than $\zeta^{-1/\alpha}$, $\lim_{\theta \to 0} S_{\beta,\xi}\left(\mathrm{PRM}\|\mathrm{PhyM} \right) \geq 1 - e^{-\theta^2 C}$, for some constant $C\geq0$.
  \item \emph{Scaling with $\lambda_t$:} For any constant $r_{\mathrm{PRM}}$ no larger than $\zeta^{-1/\alpha}$, $\lim_{\lambda_t \to \infty} S_{\beta,\xi}\left(\mathrm{PRM}\|\mathrm{PhyM} \right) \geq 1 - e^{-\lambda_t C}$ for some constant $C\geq0$.
  \item \emph{Scaling with $\lambda_o$:} For any constant $r_{\mathrm{PRM}}$ no larger than $\zeta^{-1/\alpha}$, $\lim_{\lambda_o \to 0} S_{\beta,\xi}\left(\mathrm{PRM}\|\mathrm{PhyM} \right) \geq 1 - \exp\{-C\}$ for some constant $C\geq0$, and $\lim_{\lambda_o \to \infty} S_{\beta,\xi}\left(\mathrm{PRM}\|\mathrm{PhyM} \right) \geq 1 - e^{-\lambda_{o}^{-2}D}$ for some constant $D\geq0$.
\end{itemize}
\end{result}
Due to lack of space and complexity of the analysis, we leave scaling laws of the IBM for a future publication. In~Appendix~C, we have used the Chernoff bound to bound $\Pr \left[\gamma^{\mathrm{PRM}}  < \beta \right]$, which is the first step to derive scaling laws for the IBM.

\subsection{Numerical Illustrations}
Using similar setting as in Section~\ref{sec: NumericalResult2}, Fig.~\ref{mmWave_Direc_Deterministic__lambdaT} shows the accuracy index of both IBM and PRM under Scenario~3 against $d_t$. Comparing this figure to Fig.~\ref{mmWave_Direc_Deterministic__lambdaT2}, we observe that directionality and blockage can further boost the accuracy index when we have a deterministic wireless channel. Surprisingly, \emph{the PRM is accurate enough to motivate adopting this model to analyze and design of mmWave networks instead of the PhyM, TIM, and even IBM.}
For relatively pencil-beams (e.g., $\theta=10\sim20 \degree$), which may be used in wireless backhauling applications, the accuracy of the PRM in detecting outage events is almost 1 in all our simulations. Compared to the PRM, the PhyM and IBM respectively have less than 5\% and 2\% higher accuracy in modeling the interference and detecting the outage events, but with substantially higher complexities. These complexities often result in limited (mostly intractable) mathematical analysis and little insight. More interestingly, the relative difference between the average rate of the typical link computed by the PRM and that of computed by the PhyM, namely $\mathbf{E}[\log_2(1+\gamma^{\mathrm{x}})]$ and $\mathbf{E}[\log_2(1+\gamma^{\mathrm{y}})]$ is less than 0.002\%, implying the accuracy of the simple PRM to analyze long-term performance metrics (such as throughput and delay).
\begin{figure}[!t]
  \centering
  \includegraphics[width=0.7\columnwidth]{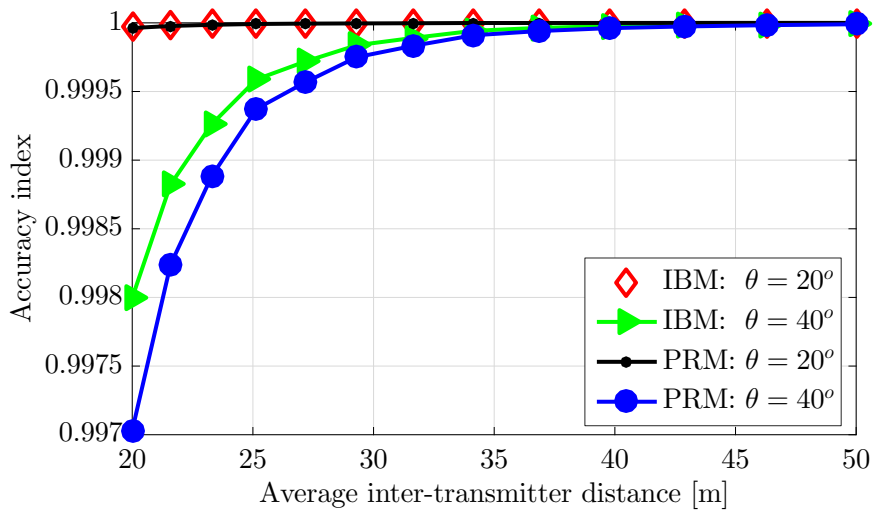}

  \caption{Accuracy of IBM and PRM under deterministic channel and directional communications. $r_{\text{PRM}} = \zeta^{-1/\alpha}$ where $\zeta$ is given in~\eqref{eq: zeta}, and $r_{\text{IBM}} = 2 r_{\text{PRM}}$. The relative difference between the average per-user rate computed by the PRM and that of computed by the PhyM is less than 0.002\%.}
  \label{mmWave_Direc_Deterministic__lambdaT}
\vspace{-7mm}
\end{figure}

Fig.~\ref{mmWave_Direc_Deterministic__lambdaT} together with Results~\ref{result: impact-of-directionality}-\ref{result: 7} support the validity of the previously proposed \emph{pseudo-wired model}~\cite{Singh2011Interference}, at least for sparse networks like mmWave mesh networks~\cite{Yaghoubi2016Mitigation}.
This highlights the importance of having quantitative (not only qualitative) insight of the accuracy of different interference models we may face in different wireless networks. Thereby, we can adopt a simple yet accurate enough model for link-level and system-level performance analysis.

So far, we have observed how we can simplify the set of dominant interferers and how much accuracy loss they entail under three network scenarios. Besides the set of interferers $\mathcal{I}$, computing the SINR expression requires modeling the wireless channel and the antenna patterns. More accurate models generally reduce tractability of the SINR expression and therefore the interference model. In the next section, we analyze the possibility of adopting simple models for the wireless channel and for the antenna pattern.

\section{Example Scenario~4: Impact of Other Components of the SINR Expression}\label{sec: FurtherDiscussions}
In this section, we analyze the accuracy loss due to simplifying wireless channel model and antenna pattern of the SINR expression. In particular, we use the proposed accuracy index to investigate the feasibility of modeling a random fading channel with a \emph{constant} value without affecting the long-term performance of the real system (with random fading). The importance of this scenario is due to that numerous studies develop protocols and optimize the network based on deterministic wireless channels, yet no study focuses on the accuracy and validity of this underlying model. In the following, we comment on what this deterministic channel gain should be to maximize its similarity to the actual random wireless channel.
We then use the proposed accuracy index to assess the impact of neglecting the reflections, assuming impenetrable obstacles, and neglecting sidelobe gain of the directional antenna on the accuracy of the resulting interference model. We consider the PhyM for both $\mathrm{x}$ and $\mathrm{y}$ throughout this section.

\subsection{Approximating a Fading Channel with a Deterministic One}\label{sec: FadingChannelDeterministic}
To design many protocols for wireless networks (such as power control, scheduling, and routing), it is often preferable to use deterministic channel gains that depend only on the distance among the transmitters and receivers~\cite{badia2008general,chen2006cross,Singh2011Interference,Singh2009Blockage,Stahlbuhk2016Topology}. In this subsection, we investigate the accuracy of approximating the fading gain between transmitter $i$ and the reference receiver ($h_i$) in $\mathrm{y}$ by a deterministic value $c_0$ in $\mathrm{x}$. After this approximation, the channel gain in $\mathrm{x}$ becomes
$g_i^{\mathrm{Ch}} = a c_0 d_{i}^{-\alpha}$, and all other parameters of $\mathrm{x}$ are identical to those of $\mathrm{y}$. For sake of simplicity, we consider omnidirectional communications without blockage, as in Section~\ref{sec: Rayleigh-noBlockage}.

\begin{figure}[!t]
  \centering
  \includegraphics[width=0.7\columnwidth]{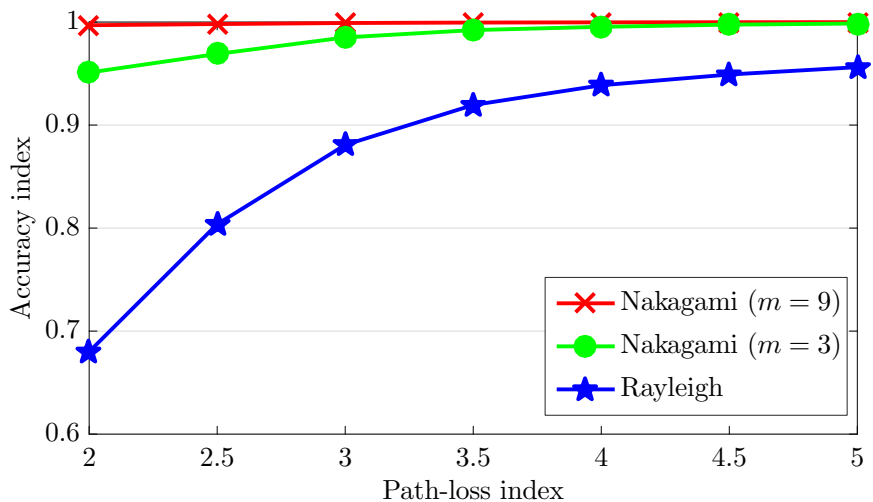}

  \caption{Impact of modeling a fading channel by a deterministic one on the accuracy of the resulting interference model ($d_t = 80$~m).}
  \label{fig: FadingDeterministic}
\vspace{-5mm}
\end{figure}
\begin{table}[!t]
  \centering
  \caption{Accuracy of the mathematical analysis when we replace fading channels with a deterministic one ($d_t = 80$~m). ``AI'' refers to our accuracy index, shown also in Fig.~\ref{fig: FadingDeterministic}. ``BC'' refers to the Bhattacharyya coefficient of the SINR distributions of $\mathrm{x}$ and $\mathrm{y}$, and ``TD'' refers to the deviation of the throughput obtained by interference model $\mathrm{x}$ from that of $\mathrm{y}$.}
  \label{table: FadingDeterministic}
{
\renewcommand{\tabcolsep}{5pt}
\renewcommand{\arraystretch}{0.85}
\begin{tabular}{|c|c|c|c|c|c|}
\hline
\multicolumn{2}{|c|}{Fading type} & $\alpha=2$ & $\alpha=3$ & $\alpha=4$ & $\alpha=5$  \\ \hline
\multirow{3}{*}{Rayleigh}  & AI & 0.68 & 0.881 & 0.939 & 0.956 \\ \cline{2-6}
& BC & 0.275 & 0.048 & 0.014 & 0.005 \\ \cline{2-6}
 & TD & 13\% & 9.3\% & 6.7\% & 4.5\%  \\ \Xhline{2\arrayrulewidth}
\multirow{3}{*}{Nakagami ($m=3$)}   & AI & 0.951 & 0.985 & 0.995 & 0.998 \\ \cline{2-6}
& BC & 0.01 & 0.004 & 0.003 & 0.002 \\ \cline{2-6}
 & TD & 5.8\% & 4.1\% & 3.2\% & 2\%  \\ \Xhline{2\arrayrulewidth}
\multirow{3}{*}{Nakagami ($m=9$)}   & AI & 0.997 & 0.9991 & 0.9996 & 0.9999 \\ \cline{2-6}
& BC & 0.001 & 0.0008 & 0.0006 & 0.0003 \\ \cline{2-6}
 & TD & 1.4\% & 1\% & 0.7\% & 0.3\% \\ \hline
\end{tabular}
}
\vspace{-7mm}
\end{table}
Using the same simulation setup as of Section~\ref{sec: Rayleigh-noBlockage}, we numerically find $c_0$ in $\mathrm{x}$ that gives the highest similarity between $\mathrm{x}$ and $\mathrm{y}$, averaged over all $\beta \in [0,10]$~dB. Fig.~\ref{fig: FadingDeterministic} shows the accuracy index, obtained by the optimal $c_0$, for Rayleigh and Nakagami fading. Moreover, we report in Table~\ref{table: FadingDeterministic} the Bhattacharyya coefficient between SINR distribution of $\mathrm{x}$ and that of $\mathrm{y}$, and also the relative difference in corresponding average throughput. From Fig.~\ref{fig: FadingDeterministic} and Table~\ref{table: FadingDeterministic}, interference model $\mathrm{x}$ (with deterministic channel) becomes more similar to $\mathrm{y}$ (with fading channel) as the path-loss index grows. This higher similarity manifests itself in higher accuracy indices, in lower Bhattacharyya coefficients, and also in lower errors in the rate analysis. Moreover, approximating a random wireless channel gain with Rayleigh fading and a small path-loss index (outdoor environment) by a constant value\footnote{We observed in our simulations that $c_0 = \mathbf{E}_h[h^{2/\alpha}] = \Gamma(1+2/\alpha)$ is roughly the optimal constant that provides the highest similarity index in Rayleigh fading channel. Notice that it is $2/\alpha$-th moment of random variable $h$.} may lead to a non-negligible inaccuracy in the final throughput analysis (up to 13\% error in our example). However, a Nakagami-$m$ fading channel with high $m$ can be well approximated by a deterministic channel gain, substantially simplifying the mathematical analysis and protocol development. The error due to this approximation will be reduced with $m$. To highlight the importance of this observation, we note that the directional communications will be largely applied in future wireless networks~\cite{boccardi2014Five}. Therefore, wireless networks with Nakagami-$m$ fading channels will play a major role in future of wireless networks. For mmWave communications, for instance, we are already using narrow beams~\cite{shokri2015mmWavecellular,Yaghoubi2016Mitigation}, which result in high $m$ in the corresponding Nakagami-$m$ fading channel. The following conjecture states how we can approximate a Nakagami-$m$ fading channel by a deterministic channel gain.

\begin{conjecture}
Consider a 2D network. Assume that the wireless channel attenuation consists of a constant attenuation at a reference distance, a distance-dependent attenuation with path-loss index $\alpha$, and a random fading $h$. If $h$ has a Nakagami-$m$ distribution with $m \geq 3$, the wireless channel can be well approximated by a deterministic LoS channel without significant drop in the accuracy of the resulting interference model or in the analysis of the ergodic performance metrics such as spectral efficiency, energy efficiency, throughput, and delay. If $h$ has Rayleigh fading distribution, replacing $h$ by its $2/\alpha$-th moment, namely $\mathbf{E}_h[h^{2/\alpha}]$, results in sufficiently accurate analysis of the ergodic performance metrics.
\end{conjecture}

In the following, we further exemplify the proposed index to investigate the accuracy drop due to simplifying other parameters of the SINR expression.

\subsection{Other Components of the SINR}\label{sec: OtherComponents}
In this subsection, we focus on mmWave networks and propose a very simple yet accurate interference model. In particular, we consider a PRM wherein we assume that \emph{i}) obstacles are impenetrable, \emph{ii}) there is no reflection, and \emph{iii}) there is no sidelobe transmissions/receptions. Although these assumptions do not generally hold in practice, we show that this simple interference model can be very accurate abstraction of real mmWave networks.
Previously, references~\cite{Singh2011Interference,Singh2009Blockage} used this interference model for performance evaluation and protocol development for mmWave networks. Therefore, the discussions of this subsection are a complementary study of those works.

We consider a random number of obstacles in the environment each with penetration loss $l_o$. The obstacles are assumed to have rectangular shape whose centers follow a spatial Poisson distribution with density $\lambda_o$ on the plane, independent of the Poisson process of the interferers. To each rectangle, we associate a random width that is independently uniformly taken from $[0,4]$~meters, a random length that is independently uniformly taken from $[0,3]$~meters, and a random orientation that is independently uniformly taken from $[0,2\pi]$. The obstacles can represent small buildings, human bodies, and cars. We independently randomly mark some obstacles to be reflectors with coefficient $r \leq 1 $. Without loss of generality, we mark the obstacles as reflectors with probability 0.1. We also assume that the links can be established either by the direct path or by a first-order reflected path. We consider a large scale path loss model at 28~GHz~\cite{Akdeniz2014MillimeterWave}, which consists of a constant attenuation, a distance dependent attenuation, and a large scale log-normal fading. Besides these attenuation sources, we consider the penetration and reflection losses. Consider path $k$ between transmitter $i$ and the reference receiver. Let $d_{ik}$ be the distance of this path (path length), $n_k$ be the number of obstacles in this path, $l_o$ be the penetration loss due to any obstacle in dB, and $l_r = -10\log(r)$ be the reflection loss in dB. Let $\mathds{1}_k$ denote an indicator function that takes 1 if path $k$ contains a reflector, otherwise 0. Then, the channel gain in $k$-th path between transmitter $i$ and the reference receiver is modelled as
\begin{equation}\label{eq: channel-gain-1}
g_{ik}^{\mathrm{Ch}} \:  [\mbox{dB}]= -61.4 - 20\log(d_{ik}) + \mathds{1}_k 10\log(r) - n_j l_{o} - X \:,
\end{equation}
where $X$ is a zero mean i.i.d. Gaussian random variable with standard deviation $5.8$~\cite{Akdeniz2014MillimeterWave}. Note that the atmospheric absorbtion is almost negligible (0.15~dB/Km) at the 28~GHz~\cite{Rangan2014Millimeter}. Moreover, changing the carrier frequency will change the parameters of channel model~\eqref{eq: channel-gain-1}, without affecting the generality of the results of this subsection. Again, we consider ideal sector antenna pattern, formulated in~\eqref{eq: SectorAntennaGains}, at all transmitters and receivers.

We consider a realistic reference physical model $\mathrm{y}$ with finite penetration loss ($l_o < \infty$), first-order reflections ($r > 0$), and non-zero antenna side lobe ($z >0$). We execute several experiments in which we change the type of the reflectors, type of obstacles, side lobe gain, operating beamwidth, and the average number of interferers and obstacles. We execute four sets of experiments. For each experiment, we compute the average accuracy index over $10^5$ random topologies and report the result in Table~\ref{table: joint-impacts}.
\begin{table}[!t]
  \centering
  \caption{Effects of assuming infinite penetration loss, no reflection, and no sidelobe gain on the accuracy of the resulting interference model. The shown parameters are for reference model $\mathrm{y}$. SINR threshold is $\beta = 5$~dB. $d_t = 1/\sqrt{\lambda_t}$ and $d_o = 1/\sqrt{\lambda_o}$.}
  \label{table: joint-impacts}
{
\renewcommand{\tabcolsep}{4.5pt}
\renewcommand{\arraystretch}{0.8}
\begin{tabular}{c!{\vrule width1pt}cccccc !{\vrule width1pt}c}
\hline \hline
Experiment & $~~l_o~~$ & $~~r~~$ & $~~z~~$ & $~~\theta~~$ & $~~d_t~~$ & $~~d_o~~$ & Accuracy \\ \hline
 1   &    10     &     0.63     &    -10    &   20$\degree$   &    50     &    20   &    0.9998        \\
 2   &    10     &     0.74     &    -10    &   40$\degree$   &    30     &    20   &    0.9992        \\
 3   &    20     &     0.9      &    -10    &   40$\degree$   &    50     &    50   &    0.9993        \\ \hline
 4   &    10     &     0.74     &    -10    &   20$\degree$   &    50     &    50   &    0.9614        \\
 5   &    20     &     0.74     &    -10    &   20$\degree$   &    30     &    50   &    0.9856        \\
 6   &    20     &     0.74     &    -10    &   20$\degree$   &    30     &    20   &    0.9588        \\ \hline
 7   &    15     &     0.74      &    -5   &   20$\degree$   &    50     &    20   &    0.9235        \\
 8   &    15     &     0.74      &    -5    &   20$\degree$   &  20     &    50    &    0.7090        \\
 9   &    15     &     0.74     &    -10   &   40$\degree$   &     30      &    20   &    0.9311        \\ \hline
 10   &    25    &     0.9      &    -10   &   10$\degree$   &    30     &    30   &    0.8810        \\
 11   &    15    &     0.63      &    -15    &   30$\degree$   &  50     &    50    &    0.9473        \\
 12   &    15    &     0.74      &    -10    &   20$\degree$   &  100     &    50    &    0.9718        \\
\hline \hline
\end{tabular}
}
\vspace{-7mm}
\end{table}

\textbf{Effects of no reflection assumption:} In the first set of experiments (1--3), we consider three materials for the reflectors: drywall with reflection coefficient 0.63, clear glass with reflection coefficient 0.74, and the tinted glass with reflection coefficient~0.9~\cite{Zhao201328Ghz}. All parameters of interference models $\mathrm{x}$ and $\mathrm{y}$ are similar (reported in the table), except that $r = 0$ in $\mathrm{x}$. From Table~\ref{table: joint-impacts}, the accuracy index is near $1$ for all scenarios. The accuracy marginally decreases with the density of the transmitters, yet it is high enough for typical transmitter densities ($d_t > 30$~m in downlink cellular networks). Increasing the operating beamwidth has similar effect as increasing the transmitter densities.

\textbf{Effects of infinite penetration loss assumption:}  In the second set of experiments (4--7), we consider different penetration losses for the obstacles. All parameters of $\mathrm{x}$ and $\mathrm{y}$ are similar (reported in the table), except that $l_o = \infty$ in $\mathrm{x}$. From the table, the accuracy index reduces with the density of the obstacles, as more obstacles correspond to higher source of errors in $\mathrm{x}$. Moreover, the assumption of impenetrable obstacles is more accurate for higher penetration loss values. Moreover, denser mmWave networks ($d_t = 30$) are less sensitive to assuming infinite penetration loss. The main reason is that densifying the network increases the probability of having interferers with LoS condition to the reference receiver. The contribution of those non-blocked interferers in the aggregated interference term dominate that of blocked interferers.

\textbf{Effects of no side lobe gains assumption:} In the third set of experiments (7--9), we investigate the impact of neglecting antenna side lobes $z$ in interference model $\mathrm{x}$. All parameters of $\mathrm{x}$ and $\mathrm{y}$ are similar (reported in the table), except that $z = 0$ in $\mathrm{x}$. Expectedly, neglecting higher $z$ lowers the accuracy of $\mathrm{x}$, and this error increases also with the number of interferers in the network. Unlike the previous parameters, neglecting side lobe gain may lead to a large deviation of $\mathrm{x}$ from $\mathrm{y}$. From the numerical results, not shown in this paper due to the space limitations, if we have either a typical dense network \footnote{Note that using scheduling, we can reduce mutual interference by controlling the number of simultaneous active transmitters. Therefore, the number of transmitters in the environment is not necessarily equal to the number of interferers; rather, it is usually much higher than that.} with $d_t = 30$ or enough side lobe suppression (at least 10~dB), we are safe to ignore side lobe gains from the interference model. Increasing the operating beamwidth increases the chance of observing an aligned interferer (which contributes in the link budget with its main lobe gain). As such interferers have dominant role in the aggregated interference term, increasing the operating beamwidth can improve the accuracy of $\mathrm{x}$.

\textbf{Joint effects of all parameters:} In the last set of experiments (10--12), we analyze joint effects of all those parameters by considering infinite penetration loss, zero reflection coefficient, and zero side lobe gain in $\mathrm{x}$. Other parameters of $\mathrm{x}$ are similar to those of $\mathrm{y}$, reported in Table~\ref{table: joint-impacts}. From the results, our simple interference model $\mathrm{x}$ is sufficiently accurate for typical mmWave network scenarios.
On the negative side, larger number of interferers magnifies the small error due to neglecting antenna side lobes. This magnified error together with other approximations leads to 12\% error in detecting outage event by the simplified interference model in Experiment~10. On the positive side, this higher transmitter density reduces the error due to both neglecting reflection and assuming impenetrable obstacles.

\section{Future Directions}\label{sec: future_directions}
Throughout this paper, we highlighted the tradeoff between the accuracy and mathematical tractability of the interference models, and exemplified the use of our accuracy index to optimize such tradeoff for different wireless network scenarios, with specific reference to mmWave networks. Although we have simplified system models of the examples to avoid unnecessary complications, our index poses no limitation to these example scenarios. We have recently used this index to assess the accuracy of a simple interference model for a mmWave cellular network~\cite{jiang2016simplified}.
Two future directions can be envisioned from this paper.

First, one may use our accuracy index to simplify the existing and develop new interference models for various network settings. In particular, illustrative examples of this paper were more suitable for ad~hoc networks, and evaluating the generality of the resulting insights is an interesting future research line. Moreover, our proposed index can be used to assess the accuracy of different blockage models like one-ball~\cite{TBai2014Blockage}, two-ball~\cite{di2014stochastic}, cone~\cite{Shokri2015Transitional}, and queue-based models~\cite{Congiu2016RelayFalback} and even develop novel accurate yet tractable models.

Second, we can extend the index itself. In this paper, we have defined the similarity index for any interference model $\mathrm{x}$ based on its ability to correctly predict the outage events; see Definition~\ref{def: IMS-index}. To generalize our approach, one may aim at measuring the similarity based on any other functions of SNIR. For example, given some alternatives for one function inside SINR (e.g., different set of interferers or different antenna models), one may use an extension of our approach to identify which of them better balances the accuracy-complexity tradeoff for a throughput/delay analysis.

\section{Conclusion}\label{sec: ConcludingRemarks}
We developed a new mathematical framework to address very fundamental questions in analysis and design of wireless networks: how accurate different interference models are and how to select the right one. We proposed a new accuracy index that quantifies the ability of any interference model in correctly predicting outage events, under any network setting. We analytically and numerically illustrated the use of our index via many example scenarios. In particular, we evaluated the accuracy of the prominent techniques that model the set of dominant interferers. We then showed that directional antenna and obstructions (basic characteristics of mmWave networks) substantially enhance the accuracy of any interference model, making the simple classical protocol model accurate enough for analysis and optimization of such networks. Furthermore, we measured the accuracy of approximating a random fading wireless channel with a deterministic channel. We conjectured that a Nakagami-$m$ fading channels with $m \geq 3$ can be well approximated by a deterministic value without introducing a significant gap in the ergodic performance metrics (e.g., throughput and delay); whereas, such gap is generally non-negligible under Rayleigh fading channels. Finally, we showed surprisingly high accuracy of a simple interference model that assumes (i) infinite penetration loss, (ii) no reflection, and (iii) no antenna side lobes in modeling a typical mmWave network where none of those assumptions hold.

\allowdisplaybreaks
\section*{Appendix~A: Deriving Components of Example Scenario~1}
\subsection{The Interference Ball Model}
Let $\mathbf{E}_{x}$ denote expectation over random variable $x$. From~\eqref{eq: general-model} and~\eqref{eq: protocol-model},
\begin{align}\label{eq: IBM-1}
\Pr\left[\gamma^{\mathrm{IBM}} < \beta \right]  &= \Pr\left[\frac{p c h_0 d_{0}^{-\alpha}}{\sum\limits_{k \in \mathcal{I} \cap \mathcal{B}(2 \pi, 0, r_{\mathrm{IBM}})} p c h_k d_{k}^{-\alpha \mathds{1}_{\overline{\mathcal{B}}(2\pi,0,a)}} + \sigma}  < \beta \right] \nonumber \\
& = 1 - \Pr\left[h_0 \geq \left( \sum\limits_{k \in \mathcal{I} \cap \mathcal{B}(2 \pi, 0, r_{\mathrm{IBM}})} h_k d_{k}^{-\alpha \mathds{1}_{\overline{\mathcal{B}}(2\pi,0,a)}} + \frac{\sigma}{pc}\right)\beta d_{0}^{\alpha} \right]  \nonumber \\
& = 1 - \mathbf{E}_{\mathcal{I},h}\left[\exp\left\{ - \left( \sum\limits_{k \in \mathcal{I} \cap \mathcal{B}(2 \pi, 0, r_{\mathrm{IBM}})} h_k d_{k}^{-\alpha \mathds{1}_{\overline{\mathcal{B}}(2\pi,0,a)}} + \frac{\sigma}{pc}\right)\beta d_{0}^{\alpha} \right\} \right]  \nonumber \\
& = 1 - \exp\left\{ \frac{- \sigma \beta d_{0}^{\alpha}}{pc} \right\}\mathbf{E}_{\mathcal{I},h}\left[\exp\left\{ -\left( \sum\limits_{k \in \mathcal{I} \cap \mathcal{B}(2 \pi, 0, r_{\mathrm{IBM}})} h_k d_{k}^{-\alpha \mathds{1}_{\overline{\mathcal{B}}(2\pi,0,a)}} \right)\beta d_{0}^{\alpha} \right\} \right]\nonumber \\
& = 1 - \exp\left\{ \frac{- \sigma \beta d_{0}^{\alpha}}{pc} \right\}\underbrace{\mathbf{E}_{\mathcal{I}}\left[\prod\limits_{k \in \mathcal{I} \cap \mathcal{B}(2 \pi, 0, r_{\mathrm{IBM}})} \mathbf{E}_h \bigg[ \exp\left\{ - \beta d_{0}^{\alpha} h_k d_{k}^{-\alpha \mathds{1}_{\overline{\mathcal{B}}(2\pi,0,a)}} \right\} \bigg] \right] }_{A} \:.
\end{align}
From probability generating functionals, we have

\begin{align}\label{eq: IBM-2}
A & = \exp\Vast\{-2 \pi  \lambda_t  \int_{0}^{\infty} \! \mathds{1}_{\mathcal{B}(2\pi,0,r_{\mathrm{IBM}})} \bigg( 1 - \mathbf{E}_h \left[e^{- \beta d_{0}^{\alpha} h r^{-\alpha \mathds{1}_{\overline{\mathcal{B}}(2\pi,0,a)}}} \right] \bigg) r \, \mathrm{d}r \Vast\} \nonumber \\
& = \exp\Vast\{-2 \pi  \lambda_t  \mathbf{E}_h \vast[\int_{0}^{\infty} \! \mathds{1}_{\mathcal{B}(2\pi,0,r_{\mathrm{IBM}})} \left( 1 - e^{- \beta d_{0}^{\alpha} h r^{-\alpha \mathds{1}_{\overline{\mathcal{B}}(2\pi,0,a)}}} \right) r \, \mathrm{d}r \vast] \Vast\} \nonumber \\
& = \exp\Vast\{-2 \pi  \lambda_t  \mathbf{E}_h \vast[\int_{0}^{a^{-}} \! \left( 1 - e^{- \beta d_{0}^{\alpha} h} \right) r \, \mathrm{d}r + \int_{a}^{r_{\mathrm{IBM}}} \! \left( 1 - e^{- \beta d_{0}^{\alpha} h r^{-\alpha}} \right) r \, \mathrm{d}r \vast] \Vast\} \nonumber \\
& \stackrel{(\star)}{=} \exp\Vast\{-\pi  \lambda_t  \mathbf{E}_h \vast[ a^2 \left( 1 - e^{- \beta d_{0}^{\alpha} h} \right) + r_{\mathrm{IBM}}^2 \left( 1 - e^{- \beta d_{0}^{\alpha} h r_{\mathrm{IBM}}^{-\alpha}} \right)
- a^2 \left( 1 - e^{- \beta d_{0}^{\alpha} h a^{-\alpha}} \right) \vast. \Vast. \nonumber \\[-4mm]
& \hspace{25mm} \Vast. \vast. + \left( \beta d_{0}^{\alpha} h \right)^{2/\alpha} \Gamma\left( 1 - \frac{2}{\alpha}, \beta d_{0}^{\alpha} h r_{\mathrm{IBM}}^{-\alpha} \right)
- \left( \beta d_{0}^{\alpha} h \right)^{2/\alpha} \Gamma\left( 1 - \frac{2}{\alpha}, \beta d_{0}^{\alpha} h a^{-\alpha} \right) \vast] \Vast\}  \:, \\[6mm]
& \nonumber
\end{align}
where $\Gamma\left(\cdot, \cdot \right)$ is the incomplete Gamma function, $(\star)$ is derived using integration by parts, and probability density function of $h$ is $f_h(x) = e^{-x}$.
To find $\Pr \left[\gamma^{\mathrm{PhyM}} < \beta \right]$, we only need to evaluate $\Pr \left[\gamma^{\mathrm{IBM}} < \beta \right]$ at $r_{\mathrm{IBM}} \to \infty$, that is,
\begin{align}\label{eq: PhyM-1}
\Pr \left[\gamma^{\mathrm{PhyM}} < \beta \right] & = \lim_{r_{\mathrm{IBM}} \to \infty} \, \Pr\left[\gamma^{\mathrm{IBM}} < \beta \right] \nonumber \\
&= 1 - \exp\Vast\{ - \frac{\sigma \beta d_{0}^{\alpha}}{pc} - \pi  \lambda_t  \mathbf{E}_h \vast[ a^2 \left( 1 - e^{- \beta d_{0}^{\alpha} h} \right) - a^2 \left( 1 - e^{- \beta d_{0}^{\alpha} h a^{-\alpha}} \right) \vast. \Vast. \nonumber \\[-4mm]
& \hspace{25mm} \Vast. \vast. + \left( \beta d_{0}^{\alpha} h \right)^{2/\alpha} \Gamma\left( 1 - \frac{2}{\alpha} \right)
- \left( \beta d_{0}^{\alpha} h \right)^{2/\alpha} \Gamma\left( 1 - \frac{2}{\alpha}, \beta d_{0}^{\alpha} h a^{-\alpha} \right) \vast] \Vast\}  \:,
\end{align}
where $\Gamma\left(\cdot \right)$ is the Gamma function.

\subsection{The Protocol Model}
Event $\gamma^{\mathrm{PRM}} \geq \beta$ implies that there is no interferer inside $\mathcal{B}(2\pi,0,r_{\mathrm{PRM}})$. Assuming $r_{\mathrm{PRM}} \geq a$ and $d_0 \geq a$, and following similar steps as in~\eqref{eq: IBM-1} and~\eqref{eq: IBM-2}, we have
\begin{align}\label{eq: PRM-2}
\Pr[\gamma^{\mathrm{PhyM}} < \beta \mid \gamma^{\mathrm{PRM}} \geq \beta]  &= 1 - \exp\Vast\{ -2 \pi  \lambda_t  \mathbf{E}_h \vast[\int_{0}^{\infty} \! \mathds{1}_{\mathcal{B}(2\pi,r_{\mathrm{PRM}},\infty)} \left( 1 - e^{- \beta d_{0}^{\alpha} h r^{-\alpha \mathds{1}_{\overline{\mathcal{B}}(2\pi,0,a)}}} \right) r \, \mathrm{d}r \vast] \Vast. \nonumber \\
& \hspace{45mm} \Vast. - \frac{\sigma \beta d_{0}^{\alpha}}{pc} \Vast\} \nonumber \\
& = 1 - \exp\Vast\{\frac{- \sigma \beta d_{0}^{\alpha}}{pc} -2 \pi  \lambda_t  \mathbf{E}_h \vast[\int_{r_{\mathrm{PRM}}}^{\infty} \! \left( 1 - e^{- \beta d_{0}^{\alpha} h r^{-\alpha}} \right) r \, \mathrm{d}r \vast] \Vast\} \nonumber \\
& \hspace{-28mm} = 1 - \exp\Vast\{ \frac{- \sigma \beta d_{0}^{\alpha}}{pc} - \pi  \lambda_t  \mathbf{E}_h \vast[ - r_{\mathrm{PRM}}^2 \left( 1 - e^{- \beta d_{0}^{\alpha} h r_{\mathrm{PRM}}^{-\alpha}} \right) + \left( \beta d_{0}^{\alpha} h \right)^{2/\alpha} \Gamma\left( 1 - \frac{2}{\alpha} \right) \vast. \Vast. \nonumber \\[-4mm]
& \hspace{45mm} \Vast. \vast.
- \left( \beta d_{0}^{\alpha} h \right)^{2/\alpha} \Gamma\left( 1 - \frac{2}{\alpha}, \beta d_{0}^{\alpha} h r_{\mathrm{PRM}}^{-\alpha} \right) \vast] \Vast\}  \:.
\end{align}

\section*{Appendix~B: Deriving Components of Example Scenario~2}
\setcounter{subsection}{0}
\subsection{The Interference Ball Model}
We have
\begin{align}\label{eq: IBM-3-1}
\Pr\left[\gamma^{\mathrm{IBM}} < \beta \right]  &= \Pr\left[\frac{p \left(\frac{2\pi}{\theta} \right)^2 c h_0 d_{0}^{-\alpha}}{\sum\limits_{k \in \mathcal{I} \cap \mathcal{B}(\theta, 0, r_{\mathrm{IBM}})} p \left(\frac{2\pi}{\theta} \right)^2 c h_k d_{k}^{-\alpha \mathds{1}_{\overline{\mathcal{B}}(2\pi,0,a)}} + \sigma}  < \beta \right] \nonumber \\
& = 1 - \mathbf{E}_{\mathcal{I},h}\left[\exp\left\{ - \left( \sum\limits_{k \in \mathcal{I} \cap \mathcal{B}(\theta, 0, r_{\mathrm{IBM}})} h_k d_{k}^{-\alpha \mathds{1}_{\overline{\mathcal{B}}(2\pi,0,a)}} + \frac{\sigma \theta^2}{4 pc\pi^2} \right)\beta d_{0}^{\alpha} \right\} \right]  \nonumber \\
& = 1 - \exp\left\{ - \frac{\sigma \theta^2 \beta d_{0}^{\alpha}}{4 pc\pi^2} \right\}\underbrace{\mathbf{E}_{\mathcal{I}}\left[\prod\limits_{k \in \mathcal{I} \cap \mathcal{B}(\theta, 0, r_{\mathrm{IBM}})} \mathbf{E}_h \bigg[ \exp\left\{ - \beta d_{0}^{\alpha} h_k d_{k}^{-\alpha \mathds{1}_{\overline{\mathcal{B}}(2\pi,0,a)}} \right\} \bigg] \right] }_{B} \:,
\end{align}
where
\begin{align}\label{eq: IBM-4-1}
\hspace{-10mm} B & = \exp\Vast\{ - \int_{0}^{\infty} \! \mathds{1}_{\mathcal{B}(\theta,0,r_{\mathrm{IBM}})} \bigg( 1 - \mathbf{E}_h \left[e^{- \beta d_{0}^{\alpha} h r^{-\alpha \mathds{1}_{\overline{\mathcal{B}}(2\pi,0,a)}}} \right] \bigg) \theta \lambda_I(r) r \, \mathrm{d}r \Vast\} \nonumber \\
& = \exp\Vast\{ - \frac{\theta^2}{2 \pi} \lambda_t \mathbf{E}_h \vast[ \int_{0}^{\infty} \! \mathds{1}_{\mathcal{B}(\theta,0,r_{\mathrm{IBM}})} \bigg( 1 - e^{- \beta d_{0}^{\alpha} h r^{-\alpha \mathds{1}_{\overline{\mathcal{B}}(2\pi,0,a)}}} \bigg) e^{-\epsilon\lambda_o r} r \, \mathrm{d}r \vast] \Vast\} \nonumber \\
& = \exp\Vast\{ - \frac{\theta^2}{2 \pi} \lambda_t   \mathbf{E}_h \vast[\int_{0}^{a^{-}} \! \left( 1 - e^{- \beta d_{0}^{\alpha} h} \right) e^{-\epsilon\lambda_o r} r \, \mathrm{d}r + \int_{a}^{r_{\mathrm{IBM}}} \! \left( 1 - e^{- \beta d_{0}^{\alpha} h r^{-\alpha}} \right) e^{-\epsilon\lambda_o r} r \, \mathrm{d}r \vast] \Vast\} \nonumber \\
& = \exp\Vast\{ - \frac{\theta^2}{2 \pi} \lambda_t   \mathbf{E}_h \vast[\left( 1 - e^{- \beta d_{0}^{\alpha} h} \right) \left( \frac{1 - \left( \epsilon \lambda_o a+ 1 \right) e^{-\epsilon \lambda_o a}}{\epsilon^2 \lambda_{o}^{2}}\right) + \int_{a}^{r_{\mathrm{IBM}}} \! \left( 1 - e^{- \beta d_{0}^{\alpha} h r^{-\alpha}} \right) e^{-\epsilon\lambda_o r} r \, \mathrm{d}r \vast] \Vast\} \:.
\end{align}
Also, we have
\begin{align}\label{eq: PhyM-New1-1}
\Pr \left[\gamma^{\mathrm{PhyM}} < \beta \right] & = \lim_{r_{\mathrm{IBM}} \to \infty} \, \Pr\left[\gamma^{\mathrm{IBM}} < \beta \right] \nonumber \\
&= 1 - \exp\Vast\{ - \frac{\sigma \theta^2 \beta d_{0}^{\alpha}}{4 pc\pi^2} - \frac{\theta^2 \lambda_t}{2 \pi \epsilon^2 \lambda_{o}^{2}} \mathbf{E}_h \vast[ 1 - e^{- \beta d_{0}^{\alpha} h} \left( 1 - \left( \epsilon \lambda_o a+ 1 \right) e^{-\epsilon \lambda_o a}\right) \vast. \Vast. \nonumber \\[-4mm]
& \hspace{70mm} \Vast. \vast. - \epsilon^2 \lambda_{o}^{2} \int_{a}^{\infty} \! e^{- \beta d_{0}^{\alpha} h r^{-\alpha} -\epsilon\lambda_o r} r \, \mathrm{d}r \vast] \Vast\}  \:.
\end{align}

\subsection{The Protocol Model}
We have
\begin{align}
\hspace{-5mm} \Pr[\gamma^{\mathrm{PhyM}} < \beta \mid \gamma^{\mathrm{PRM}} \geq \beta]  &= 1 - \exp\Vast\{- \mathbf{E}_h \vast[\theta \int_{0}^{\infty} \! \mathds{1}_{\mathcal{B}(\theta,r_{\mathrm{PRM}},\infty)} \left( 1 - e^{- \beta d_{0}^{\alpha} h r^{-\alpha \mathds{1}_{\overline{\mathcal{B}}(2\pi,0,a)}}} \right) \lambda_I(r) r \, \mathrm{d}r \vast] \Vast. \nonumber \\
& \hspace{60mm} - \frac{\sigma \theta^2 \beta d_{0}^{\alpha}}{4 pc\pi^2}  \Vast\} \nonumber \\
& = 1 - \exp\Vast\{- \frac{\sigma \theta^2 \beta d_{0}^{\alpha}}{4 pc\pi^2} - \frac{\theta^2 \lambda_t}{2 \pi }  \mathbf{E}_h \vast[\int_{r_{\mathrm{PRM}}}^{\infty} \! \left( 1 - e^{- \beta d_{0}^{\alpha} h r^{-\alpha}} \right) e^{-\epsilon\lambda_o r} r \, \mathrm{d}r \vast] \Vast\} \:.
\end{align}

\section*{Appendix~C: Deriving Bounds for Example Scenario~3}
\setcounter{subsection}{0}
In this appendix, we derive bounds on the miss-detection probability in Example Scenario~3.
\subsection{The Interference Ball Model}
To derive an upper bound on the miss-detection probability, we substitute a lower bound of
$\Pr \left[\gamma^{\mathrm{IBM}} < \beta \right]$ and upper bound of $\Pr \left[\gamma^{\mathrm{PhyM}} < \beta \right]$ into~\eqref{eq: miss-detection-uWave-Rayleigh-1} of the main manuscript. Recall the definition of $\zeta$ in \eqref{eq: zeta}. For any real positive $\tau$,
\begin{align}\label{eq: Chernoff-bound-1}
\Pr \left[\gamma^{\mathrm{IBM}} < \beta \right]  & \stackrel{(\star)}{=} \Pr \left[ \sum_{k \in \mathcal{I}} d_{k}^{-\alpha  \mathds{1}_{\overline{\mathcal{B}}(2\pi,0,a)}} \mathds{1}_{\mathcal{B}(\theta,0,r_{\mathrm{IBM}})} > \zeta \right] =
1 - \Pr \left[ \sum_{k \in \mathcal{I}} d_{k}^{-\alpha  \mathds{1}_{\overline{\mathcal{B}}(2\pi,0,a)}} \mathds{1}_{\mathcal{B}(\theta,0,r_{\mathrm{IBM}})} \leq \zeta \right] \nonumber \\
& \stackrel{(\star \star)}{\geq} 1 - \inf\limits_{\tau > 0} \: e^{\tau\zeta}~ \underbrace{\mathbf{E}_{\mathcal{I}} \vast[\exp\left\{-\tau \sum_{k \in \mathcal{I}} d_{k}^{-\alpha  \mathds{1}_{\overline{\mathcal{B}}(2\pi,0,a)}} \mathds{1}_{\mathcal{B}(\theta,0,r_{\mathrm{IBM}})}\right\} \vast]}_{C}  \:.
\end{align}
where $(\star)$ is due to~\eqref{eq: protocol-model} in the main manuscript, and $(\star \star)$ follows from the Chernoff bound and the probability generating functionals.
\begin{align}\label{eq: Cambell-theorem-1}
\hspace{-5mm}
C & = \exp\vast\{ - \theta \int_{0}^{r_{\mathrm{IBM}}} \! \left( 1 - e^{-\tau r^{-\alpha  \mathds{1}_{\overline{\mathcal{B}}(2\pi,0,a)}} } \right) \lambda_I (r) r \, \mathrm{d}r \vast\} \nonumber \\
&= \exp\Vast\{ - \frac{\theta^2}{2\pi}\lambda_t \vast( \int_{0}^{a^{-}} \! \left( 1 - e^{-\tau} \right) r e^{-\epsilon\lambda_o r} \, \mathrm{d}r + \int_{a}^{r_{\mathrm{IBM}}} \! \left( 1 - e^{-\tau r^{-\alpha}} \right) r e^{-\epsilon\lambda_o r} \, \mathrm{d}r \vast) \Vast\} \nonumber \\
& = \exp \Vast\{ - \frac{ \theta^2 \lambda_t}{2\pi } \vast( \frac{1 - e^{-\tau} + \left( 1 + \epsilon\lambda_o a\right)e^{-\epsilon\lambda_o a - \tau } - \left( 1 + \epsilon\lambda_o r_{\mathrm{IBM}} \right)e^{-\epsilon\lambda_o r_{\mathrm{IBM}}}}{\epsilon^2 \lambda_{o}^{2}} \vast. \Vast. \nonumber \\
 & \hspace{90mm} - \int_{a}^{r_{\mathrm{IBM}}} \! r e^{-\epsilon\lambda_o r -\tau r^{-\alpha}} \, \mathrm{d}r \vast) \Vast\} \:.
\end{align}
Using similar technique, we use the Chernoff bound to find exponentially decreasing bound on the tail distribution of $\gamma^{\mathrm{PhyM}}$ as
\begin{align}\label{eq: Chernoff-bound-2}
\hspace{-5mm} \Pr \left[\gamma^{\mathrm{PhyM}} < \beta \right]  & = \Pr \left[ \sum_{k \in \mathcal{I}} d_{k}^{-\alpha  \mathds{1}_{\overline{\mathcal{B}}(2\pi,0,a)}} \mathds{1}_{\mathcal{B}(\theta,0,\infty)} > \zeta \right] \nonumber \\
& \leq \inf\limits_{\tau > 0} \: e^{-\tau\zeta}~\mathbf{E}_{\mathcal{I}} \vast[\exp\left\{\tau \sum_{k \in \mathcal{I}} d_{k}^{-\alpha  \mathds{1}_{\overline{\mathcal{B}}(2\pi,0,a)}} \mathds{1}_{\mathcal{B}(\theta,0,\infty)}\right\} \vast] \nonumber \\
& =  \inf\limits_{\tau > 0} \: \exp \Vast\{ -\tau \zeta - \frac{ \theta^2 \lambda_t}{2\pi } \vast( \frac{1 - e^{\tau} + \left( 1 + \epsilon\lambda_o a\right)e^{-\epsilon\lambda_o a+\tau}}{\epsilon^2 \lambda_{o}^{2}} -  \int_{a}^{\infty} \! r e^{-\epsilon\lambda_o r +\tau r^{-\alpha}} \, \mathrm{d}r \vast) \Vast\} \:.
\end{align}
Note that bounds~\eqref{eq: Chernoff-bound-1} and~\eqref{eq: Chernoff-bound-2} are derived using the Chernoff bound. However, easier but looser bounds can be found using Markov and Chebyshev inequalities. These bounds can be readily derived by direct application of the Campbell's Theorem~\cite{Haenggi2013Stochastic}.

\subsection{The Protocol Model}
We can also find bounds and scaling laws for the accuracy index of the PRM in Example~Scenario~3. To this end, we use the following proposition:
\begin{prop}
For $\xi = \Pr \left[ \gamma^{\mathrm{PhyM}} \geq \beta \right]$, we have
\begin{equation}\label{S-IBM-example3-bound-t}
\max\left(\Pr \left[\gamma^{\mathrm{IBM}}  < \beta \right] , \Pr \left[\gamma^{\mathrm{PhyM}} \geq \beta \right]\right) \leq S_{\beta,\xi}\left(\mathrm{IBM}\|\mathrm{PhyM} \right) \leq 1 \:.
\end{equation}
Also, for any $0 < r_{\mathrm{PRM}} \leq \zeta^{-1/\alpha}$ where $\zeta$ is defined in \eqref{eq: zeta} of the manuscript we have
\begin{equation}\label{eq: S-PRM-example3-bound-t}
\max\left(\Pr \left[\gamma^{\mathrm{PRM}}  < \beta \right] , \Pr \left[\gamma^{\mathrm{PhyM}} \geq \beta \right]\right) \leq S_{\beta,\xi}\left(\mathrm{PRM}\|\mathrm{PhyM} \right) \leq 1 \:,
\end{equation}
where
\begin{equation*}
\Pr \left[\gamma^{\mathrm{PRM}}  < \beta \right] = 1 - \exp\left\{- \frac{\theta^2 \lambda_t}{2\pi \epsilon^{2} \lambda_{o}^{2}} \Big( 1 - \left( 1 + \epsilon \lambda_o r_{\mathrm{PRM}} \right)e^{- \epsilon \lambda_o r_{\mathrm{PRM}}}\Big) \right\} \:.
\end{equation*}
\end{prop}

\begin{IEEEproof}
For the IBM, the upper bound is trivial. To derive the lower bound, from \eqref{eq: Definition-IMSindex} of the manuscript we have
\begin{align}\label{eq: misdetection-example3-bound-1}
S_{\beta,\xi}\left(\mathrm{IBM}\|\mathrm{PhyM} \right) &  = 1 - \xi p_{\mathrm{fa}}^{{\mathrm{IBM}} \mid {\mathrm{PhyM}}} - \left( 1 - \xi\right) p_{\mathrm{md}}^{{\mathrm{IBM}} \mid {\mathrm{PhyM}}} \nonumber \\
& \stackrel{(\star)}{=} 1 - \left( 1 - \xi\right) p_{\mathrm{md}}^{{\mathrm{IBM}} \mid {\mathrm{PhyM}}} \nonumber\\
& \stackrel{(\star\star)}{=} \Pr \left[\gamma^{\mathrm{PhyM}}  \geq \beta \right] + \Pr \left[\gamma^{\mathrm{IBM}}  < \beta \right]\:,
\end{align}
where $(\star)$ is because $p_{\mathrm{fa}}^{{\mathrm{IBM}} \mid {\mathrm{PhyM}}} = 0$ for any $r_{\mathrm{IBM}} \geq 0$, and $(\star\star)$ is due to \eqref{eq: miss-detection-uWave-Rayleigh-1} of the manuscript. Then,~\eqref{S-IBM-example3-bound-t} follows.

To derive the lower bound of the PRM, we first note that $\Pr \left[\gamma^{\mathrm{PhyM}} < \beta \mid \gamma^{\mathrm{PRM}} \geq \beta \right] \leq \Pr \left[\gamma^{\mathrm{PhyM}} < \beta \right]$ for any ${r_{\mathrm{PRM}} > 0}$. Therefore, from \eqref{eq: misdetection-prop-2} of the manuscript,
\begin{align}\label{eq: misdetection-example3-bound}
p_{\mathrm{md}}^{{\mathrm{PRM}} \mid {\mathrm{PhyM}}} & \leq 1 - \Pr \left[\gamma^{\mathrm{PRM}} < \beta \right] =  \exp\left\{- \frac{\theta^2 \lambda_t}{2\pi \epsilon^{2} \lambda_{o}^{2}} \Big( 1 - \left( 1 + \epsilon \lambda_o r_{\mathrm{PRM}} \right)e^{- \epsilon \lambda_o r_{\mathrm{PRM}}}\Big) \right\}\:.
\end{align}
Then, from~\eqref{eq: Definition-IMSindex} we have
\begin{align}\label{eq: misdetection-example3-bound-1}
S_{\beta,\xi}\left(\mathrm{PRM}\|\mathrm{PhyM} \right) &  = 1 - \xi p_{\mathrm{fa}}^{{\mathrm{PRM}} \mid {\mathrm{PhyM}}} - \left( 1 - \xi\right) p_{\mathrm{md}}^{{\mathrm{PRM}} \mid {\mathrm{PhyM}}} \nonumber \\
& \stackrel{(\star)}{=} 1 - \left( 1 - \xi\right) p_{\mathrm{md}}^{{\mathrm{PRM}} \mid {\mathrm{PhyM}}} \nonumber\\
& \geq 1 - p_{\mathrm{md}}^{{\mathrm{PRM}} \mid {\mathrm{PhyM}}} \nonumber \\
& \stackrel{(\star\star)}{\geq} \Pr \left[\gamma^{\mathrm{PRM}}  < \beta \right] = 1 - \exp\left\{- \frac{\theta^2 \lambda_t}{2\pi \epsilon^{2} \lambda_{o}^{2}} \Big( 1 - \left( 1 + \epsilon \lambda_o r_{\mathrm{PRM}} \right)e^{- \epsilon \lambda_o r_{\mathrm{PRM}}}\Big) \right\} \:,
\end{align}
where $(\star)$ is because $p_{\mathrm{fa}}^{{\mathrm{PRM}} \mid {\mathrm{PhyM}}} = 0$ for any $r_{\mathrm{PRM}} \leq \zeta^{-1/\alpha}$ where $\zeta$ is defined in Eqn.~(27) of the manuscript (see Result~7 of the manuscript), and $(\star\star)$ is due to Eqn~\eqref{eq: misdetection-example3-bound}. Moreover,
\begin{align}\label{eq: misdetection-example3-bound-2}
S_{\beta,\xi}\left(\mathrm{PRM}\|\mathrm{PhyM} \right) & = 1 - \left( 1 - \xi\right) p_{\mathrm{md}}^{{\mathrm{PRM}} \mid {\mathrm{PhyM}}} \nonumber\\
& \stackrel{(\star)}{=} 1 - \Pr \left[\gamma^{\mathrm{PRM}} \geq \beta \right] \Pr \left[\gamma^{\mathrm{PhyM}} < \beta \mid \gamma^{\mathrm{PRM}} \geq \beta \right] \nonumber \\
& \geq 1 - \Pr \left[\gamma^{\mathrm{PhyM}} < \beta \mid \gamma^{\mathrm{PRM}} \geq \beta \right] \nonumber \\
& \geq 1 - \Pr \left[\gamma^{\mathrm{PhyM}} < \beta \right] \:,
\end{align}
where $(\star)$ is from~\eqref{eq: misdetection-prop-2}. Combining~\eqref{eq: misdetection-example3-bound-1} and~\eqref{eq: misdetection-example3-bound-2} results in~\eqref{eq: S-PRM-example3-bound-t}.
\end{IEEEproof}

\bibliographystyle{IEEEtran}
\bibliography{References}

\end{document}